\documentclass{jfm}
\usepackage{graphicx}
\usepackage{float}

\usepackage{xpatch}

\usepackage{amsmath} 
\usepackage{scalerel}
\usepackage{tikz}
\usetikzlibrary{svg.path}

\definecolor{orcidlogocol}{HTML}{A6CE39}
\tikzset{
  orcidlogo/.pic={
    \fill[orcidlogocol] svg{M256,128c0,70.7-57.3,128-128,128C57.3,256,0,198.7,0,128C0,57.3,57.3,0,128,0C198.7,0,256,57.3,256,128z};
    \fill[white] svg{M86.3,186.2H70.9V79.1h15.4v48.4V186.2z}
                 svg{M108.9,79.1h41.6c39.6,0,57,28.3,57,53.6c0,27.5-21.5,53.6-56.8,53.6h-41.8V79.1z M124.3,172.4h24.5c34.9,0,42.9-26.5,42.9-39.7c0-21.5-13.7-39.7-43.7-39.7h-23.7V172.4z}
                 svg{M88.7,56.8c0,5.5-4.5,10.1-10.1,10.1c-5.6,0-10.1-4.6-10.1-10.1c0-5.6,4.5-10.1,10.1-10.1C84.2,46.7,88.7,51.3,88.7,56.8z};
  }
}

\newcommand\orcidicon[1]{\href{https://orcid.org/#1}{\mbox{\scalerel*{
\begin{tikzpicture}[yscale=-1,transform shape]
\pic{orcidlogo};
\end{tikzpicture}
}{|}}}}
\usepackage[colorlinks=true, allcolors=blue]{hyperref}

\usepackage{epstopdf, epsfig}
\usepackage{xcolor}

\shorttitle{Reconstruction of irregular flow dynamics from sparse measurements}
\shortauthor{F. Savarino, G. Papadakis}

\title{Reconstruction of irregular flow dynamics around two square cylinders from sparse measurements using a data-driven algorithm}

\author{Flavio Savarino,
        George Papadakis\corresp{\email{g.papadakis@imperial.ac.uk}}}

\affiliation{Department of Aeronautics, Imperial College London, Exhibition Rd, London SW7 2AZ, UK}

\begin{document}

\maketitle

\begin{abstract}
We propose a data-driven algorithm for reconstructing the irregular, chaotic flow dynamics around two side-by-side square cylinders from sparse,  time-resolved, velocity measurements in the wake. We use Proper Orthogonal Decomposition (POD) to reduce the dimensionality of the problem and then explore two different reconstruction approaches: in the first approach, we use the subspace system identification algorithm \texttt{n4sid} to extract a linear dynamical model directly from the data (including the modelling and measurement error covariance matrices) and then employ Kalman filter theory to synthesize a linearly optimal estimator. In the second approach, the estimator matrices are directly identified using \texttt{n4sid}. A systematic study reveals that the first strategy outperforms the second in terms of reconstruction accuracy, robustness and computational efficiency. We also consider the problem of sensor placement. A greedy approach based on the QR pivoting algorithm is compared against sensors placed at the POD mode peaks; we show that the former approach is more accurate in recovering the flow characteristics away from the cylinders. We demonstrate that a linear dynamic model with a sufficiently large number of states and relatively few measurements, can recover accurately complex flow features, such as the interaction of the irregular flapping motion of the jet emanating from the gap with the vortices shed from the cylinders as well as the convoluted patterns downstream arising from the amalgamation of the individual wakes. The proposed methodology is entirely data-driven, does not have tunable parameters, and the resulting matrices are unique (to within a linear coordinate transformation of the state vector). The method can be applied directly to either experimental or computational data. 
\end{abstract}


\section{Introduction}\label{sec:introduction}
Flow estimation from limited measurements has many applications, for example in active flow control (combination with actuators to achieve an objective, such as drag reduction), in cardiovascular medicine (extraction of blood flow patterns from noisy imaging data), in environmental engineering (prediction of pollutant dispersion), etc. The literature on the subject is vast and below we present only a few key approaches to set the context for the present work. More details can be found in \cite{Brunton_Noack_2015, Sipp_Schmid_2016,Amaral, Guastoni_et_al_2021, Callaham_et_al_2019} and references therein.

The linear stochastic estimation (LSE), see \cite{Adrian_1979, adrian_moin_1988} was the first method to approximate the conditional velocity average at one point in the flow using unconditional statistics at another point. This was achieved by establishing a linear relationship between the two instantaneous quantities and computing the coefficient matrix using the two-point correlation tensor. The original method was extended to include quadratic terms and time-delays between the two quantities, \cite{guezennec89}. It has been also coupled with Proper Orthogonal Decomposition (POD),  \citet{Holmes2012TurbulenceSymmetry}, and used to infer the POD coefficients directly from pressure or velocity measurements. 

The key idea of stochastic estimation is to establish a relationship (linear or quadratic) between the measured quantity (or input) and the quantity one wants to estimate (output). The coefficient matrices are then obtained by solving a least squares problem. Other approaches that minimise the $\mathcal{L}_1$ norm (instead of the $\mathcal{L}_2$ norm of standard least squares) and lead to sparse representations have been also proposed, see \cite{Callaham_et_al_2019}.  Recently, much more general input/output mappings have been found using a variety of machine learning techniques, such as convolutional neural networks, Long Short-Term Memory (LSTM) networks, see \cite{Fukami, Carter, Giannopoulos, Guastoni_et_al_2021, Nair_Goza_2020, Kim_kim_won_lee_2021,  Rozon_Breitsamter_2021}. Although very successful reconstructions have been reported, neural network approaches require careful tuning of several parameters, such as the number of layers, time delays etc. 

Another approach, known as dynamic estimation, utilises an approximate linear dynamical model of the flow that links the input and the output. The model contains unknown uncertainties, either due to errors in representing the real flow or due to measurement noise. An optimisation problem is solved resulting in the well-known Kalman filter, see for example \cite{Kailath_Hassibi_Sayed_2000, Anderson_and_moore}. The approximate model can be obtained directly from the linearised Navier-Stokes equations and can be formulated either in the time or frequency domains. Recent estimation work has employed resolvent-based models that are formulated in the frequency domain and result in non-causal estimators, see for example \cite{Symon,Amaral,Martini_Cavalieri_Jordan_Towne_Lesshafft_2020}, and special treatment is needed to recover causality \cite{Martini_Jung_Cavalieri_Jordan_Towne_2022}. For models in the time domain, refer to \cite{Oehler_Illingworth_2018, Juniper} among others. In order to account for the forcing of finite size perturbations around the mean velocity profile, an eddy viscosity term is usually employed. However, the expression for this term is known only in simple geometries (such as flow between parallel flat plates), and is difficult to estimate in more general flow settings.

The underlying model can be also obtained from data. For example in \cite{Tu_et_al_2013} the model comprised a simple  analytic oscillatory component describing the vortex shedding activity (the frequency was obtained from the data) and a stochastic component describing the dynamics of all the other POD modes. Non-linear POD models have also been used in conjunction with particle filter and ensemble Kalman filter, see \cite{Kikuchi_et_al_2015}. 

It is also  possible to obtain the matrices of a dynamic estimator directly from data. \cite{guzman2014dynamic} combined POD with  system identification to derive a linear dynamic estimator of infinitesimal perturbations (around a laminar base flow profile) from measurements at a single point in a two-dimensional boundary layer flow.  This idea was later extended to estimate finite size fluctuations around a time-average velocity in two and three dimensional, laminar and transitional, flows \cite{Inigo_Sodar_Papadakis_2019,Mikhaylov_et_al_2021}. Non-linear system identification methods also exist. In  \citet{Loiseau} the application of the sparse identification of nonlinear dynamics (SINDy) algorithm to the laminar shedding instability of the circular cylinder wake led to the construction of a full-state estimator from sparse sensor measurements. 

In the present work we consider linear dynamical models. Such models derived directly from the linearised Navier-Stokes equations offer good reconstruction quality but require a good approximation of the forcing due to non-linear terms; this is achieved either using a turbulence model or providing directly the true spatio–temporal statistics of the forcing. On the other hand, the performance of linear models derived directly from data has been far less explored. In particular, the fact that finite amplitude perturbations about  the time-average are used to derive the linear model requires careful interpretation from the physical point of view.  When the model is obtained directly from the linearised Navier-Stokes equations, the physical interpretation is clear, but when obtained directly from data, it is much less so. 

In the present paper, we derive a data-driven model for the two-dimensional flow past two side-by-side square cylinders. The vortex shedding mechanism behind the cylinders is determined by the Reynolds number ($Re$) and the gap ($g$) between the cylinders. The combination of $Re=200$ and gap length equal to the cylinder side results in a highly irregular vortex shedding with multiple nonlinear interactions between spatio-temporal scales \citep{Ma,Shun}. We first employ POD to reduce the dimensionality of the system and then apply a system identification algorithm to find the underlying linear dynamical system that governs the evolution of the time coefficients. We argue that the system identification algorithm  implicitly extracts an eddy-viscosity matrix directly from the data, thus effectively approximating the non-linear terms in the underlying system. Modelling errors are included in a noise term, and the algorithm also provides its covariance matrix. Furthermore, because the optimisation problem solved is convex, the computed coefficient matrices are unique (to within a linear transformation of the state vector) and there are no tunable parameters. 

The strategic choice of sensor locations is important for accurate reconstruction \citep{Inigo_Sodar_Papadakis_2019, Mikhaylov_et_al_2021, Loiseau}. Previous work by \citet{Yildirim} suggests placing the sensors at the POD mode peaks. In the present paper we follow a different approach that exploits the availability of the identified linear dynamical system. More specifically, in order to find good sensor locations, we apply the QR pivoting algorithm to a matrix that involves the output controllability Gramian, and we compare the performance of QR and POD sensors. 

The paper is structured as follows: the flow configuration and main characteristics are presented in \S\ref{sec:Flow configuration and computational details} and \S\ref{sec:Analysis of flow characteristics} respectively. Section  \S\ref{sec:POD} describes the POD algorithm and the dominant modes of the flow. The system identification and optimal estimation methods using the \texttt{n4sid} algorithm are explained in \S\ref{sec:Flow reconstruction using system identification and optimal estimation theory}, while the sparse sensor placement algorithm is derived in \S\ref{sec:Sparse sensor placement}. Section  \S\ref{sec:Performance assessment of dynamic estimation} presents a systematic performance evaluation of the designed estimators. Main conclusions are summarised in  \S\ref{sec:conclusions}.


\section{Flow configuration and computational details}\label{sec:Flow configuration and computational details}
We consider the two-dimensional, incompressible  flow around two square cylinders of side $D$, separated by gap $g$ in the cross-flow direction. The flow is governed by the momentum and continuity equations,
\begin{subequations}
\begin{equation}
  \frac{\partial \boldsymbol{u}}{\partial t} + \boldsymbol{u}\cdot \nabla \boldsymbol{u} = - \nabla p + Re^{-1} \: \nabla^2 \boldsymbol{u},
  \label{subeq:momentum}
\end{equation}
\begin{equation}
  \nabla \cdot \boldsymbol{u} = 0,
  \label{subeq:incompressibility}
\end{equation} 
  \label{eq:NS}
\end{subequations}
\noindent where $\boldsymbol{u}$ and $p$ denote the non-dimensional velocity vector and pressure respectively. The reference quantity for distances is $D$, for velocities the flow speed away from the cylinders, $U_{\infty}$, and for pressure $\rho U_{\infty}^2$, where $\rho$ is the fluid density. The Reynolds number, defined as  $Re=U_{\infty}D/\nu$, where $\nu$ is the kinematic viscosity, is equal to 200, and the gap ratio  $g^{*}=\frac{g}{D}=1$. In the following, $\overline{\phi}$ denotes the time-average of the general variable $\phi$ and $\phi'$ the fluctuation, i.e.\ $\phi=\overline{\phi}+\phi'$. 

\begin{figure}[b!]
  \centerline{\includegraphics[width=0.8\textwidth]{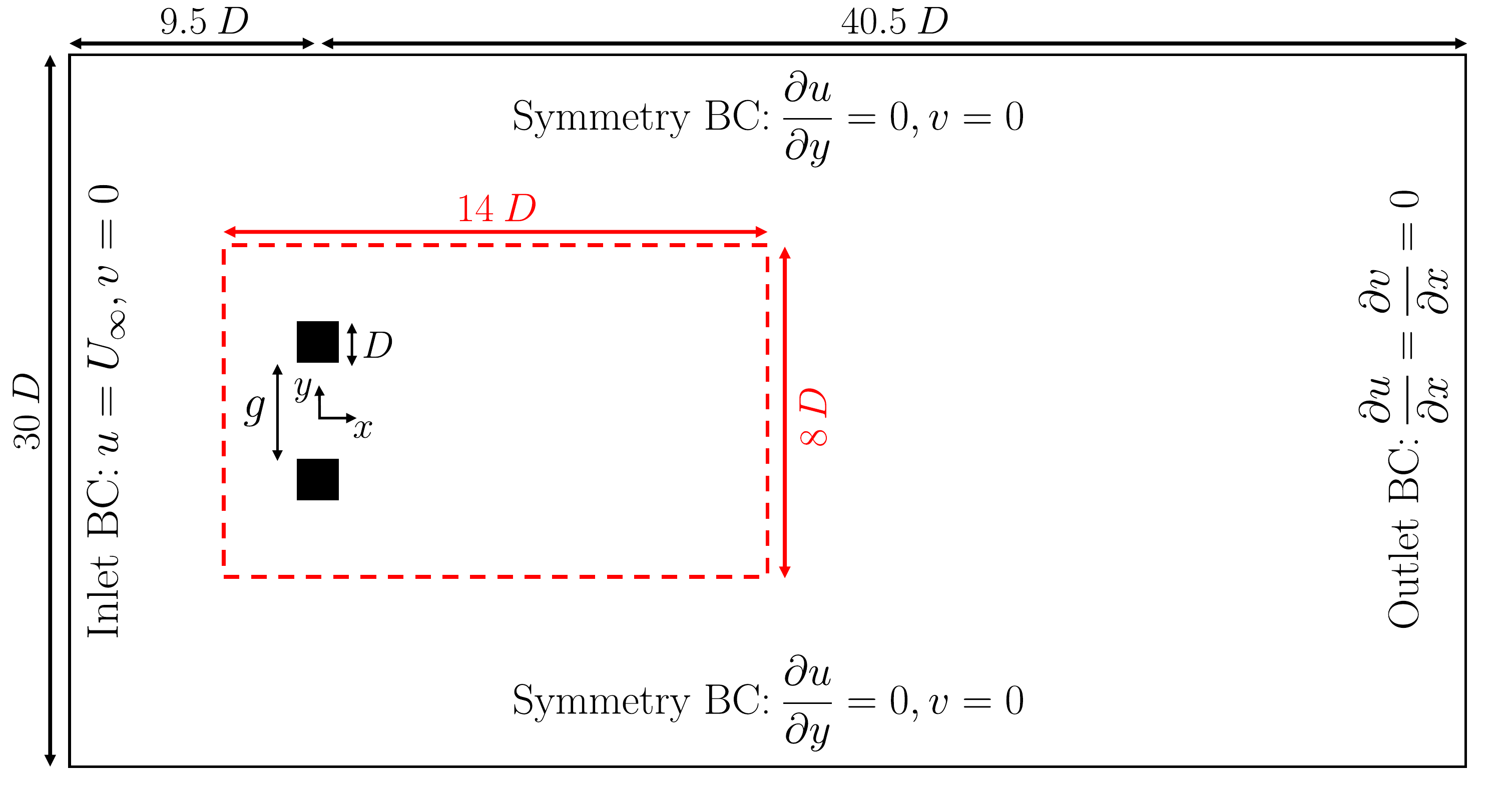}}
  \caption{Flow configuration and boundary conditions. The red dashed line indicates the boundary of the domain where the velocity field is extracted for further analysis. The figure is not to scale.}
\label{fig:geometry}
\end{figure}

The geometry, computational domain, and boundary conditions are shown in figure \ref{fig:geometry}. The origin of the coordinate system is fixed at the centre of the gap and midway along the cylinder side length. The streamwise direction is denoted as $x$ and the cross-stream direction as $y$; the corresponding velocity components are $u$ and $v$ respectively, i.e.\ $\boldsymbol{u}=(u,v)$. The far field boundaries in the streamwise direction are located 9.5$D$ upstream and 40.5$D$ downstream of the origin. In the cross-stream direction, the boundaries are placed 15$D$ above and below the centerline. Uniform horizontal velocity is prescribed at the inlet, zero-gradient at the outlet, symmetry at the top and bottom boundaries, and  no-slip at the  walls.  Velocity snapshots are recorded in a smaller domain with size $14D \times 8D$ for further analysis; the domain is marked by the red-dashed line in figure \ref{fig:geometry}.

\begin{figure}[!ht]
  \centerline{\includegraphics[width=0.65\textwidth]{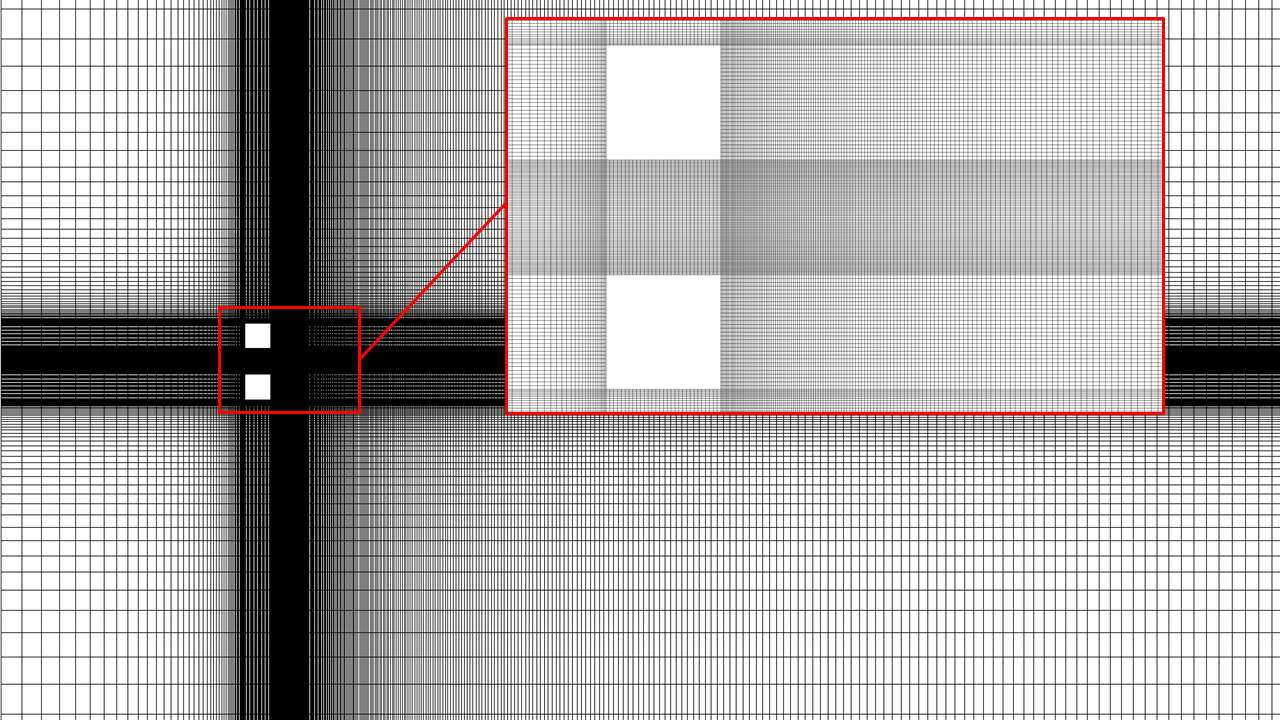}}
  \caption{Global view of the Cartesian mesh with 89,400 cells and zoomed-in view close to the two cylinders.}
\label{fig:mesh}
\end{figure}

A Cartesian mesh with grid lines densely clustered around the cylinders is employed, as shown in figure \ref{fig:mesh}. The equations are discretised using the finite volume method. A second order central scheme is employed for the spatial discretisation of the convective, viscous and pressure terms. For time advancement, a second order implicit backward differencing scheme with time step $\Delta t = 0.01$ is used. The PISO algorithm is employed to enforce the incompressibility condition. The equations are solved numerically using Ansys Fluent \citep{PISO}. Simulations were performed with two meshes; details are summarised in table \ref{tab:mesh convergence}. The mean and rms values of the aerodynamic coefficients for the top cylinder are very close, indicating that the results are almost grid independent even with the coarser mesh. The statistics were obtained from time-averaging over $2,200$ time units. This rather long simulation time is justified by the presence of low frequency flow features, which are described in more detail in the following section. The velocity dataset used for subsequent analysis was extracted from the finer mesh simulation.

\makeatletter
\let\@float@original\@float
\xpatchcmd{\@float}{\csname fps@#1\endcsname}{h!}{}{}
\makeatother

\begin{table}
  \begin{center}
\def~{\hphantom{0}}
  \begin{tabular}{ccccccc}
      \textit{Global number} & \textit{Cells across}  & \textit{Thickness of cell} & $\overline{C_D}$ & $\overline{C_L}$ & $\left(C'_D\right)_{rms}$ & $\left(C'_L\right)_{rms}$ \\
      \textit{of cells} & \textit{cylinder side} & \textit{next to the wall} & $ $ & $ $ & $ $ & $ $ \\ [3pt]
      ~62,400 & 20 & ~$0.02D$ & 2.046 & 0.203 & 0.351 & 0.745 \\
      ~89,400 & 30 & ~$0.01D$ & 2.034 & 0.204 & 0.354 & 0.740 \\
  \end{tabular}
  \caption{Mesh details and aerodynamic coefficients (mean and rms) for the top cylinder.}
  \label{tab:mesh convergence}
  \end{center}
\end{table}

\begin{figure}
    \centering
    \includegraphics[width=0.65\textwidth]{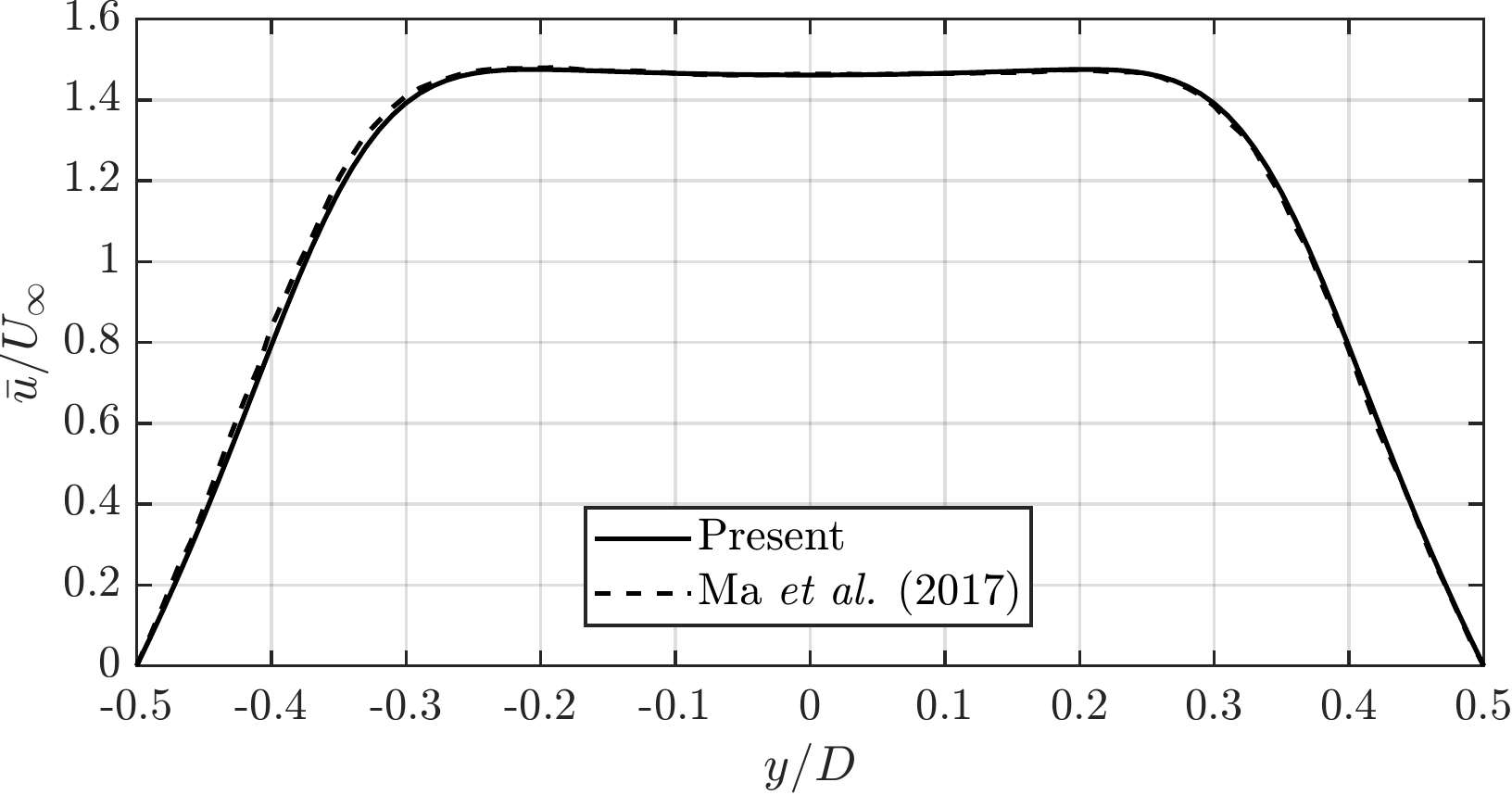}
    \caption{Normalised streamwise mean velocity profile, $\bar{u}/U_\infty$, across the gap between the cylinders. The present profile (solid line) is compared with that of \citet{Ma} (dashed line).}
    \label{fig:mean x-velocity gap}
\end{figure}

The present results are validated against the extensive numerical work of \citet{Ma}. In figure \ref{fig:mean x-velocity gap} we compare the horizontal velocity profile $\bar{u}(y)$ across the gap between the cylinders with that of \citet{Ma}; the two profiles match very well. The average velocity across the gap, 
${\bar{u}_{gap}}= \frac{1}{g^*}\int_{-0.5g^*}^{0.5g^*} {\bar{u}} \: \textup{d}y = 1.182$ matches with the reported value of 1.188. Similarly, the predicted mean drag coefficient for the top cylinder $\overline{C_D}=2.034$ agrees with the value of $2.03$, while the predicted lift coefficient $\overline{C_L}=0.204$ is also close to $0.19$. The bottom cylinder experiences the same mean drag coefficient as the top, but the lift acts in the opposite direction, because the two cylinders repel each other for the small gap considered, see \citet{Burattini}. 

\section{Analysis of the flow characteristics}\label{sec:Analysis of flow characteristics}
\begin{figure}[ht]
  \centerline{\includegraphics[width=\textwidth]{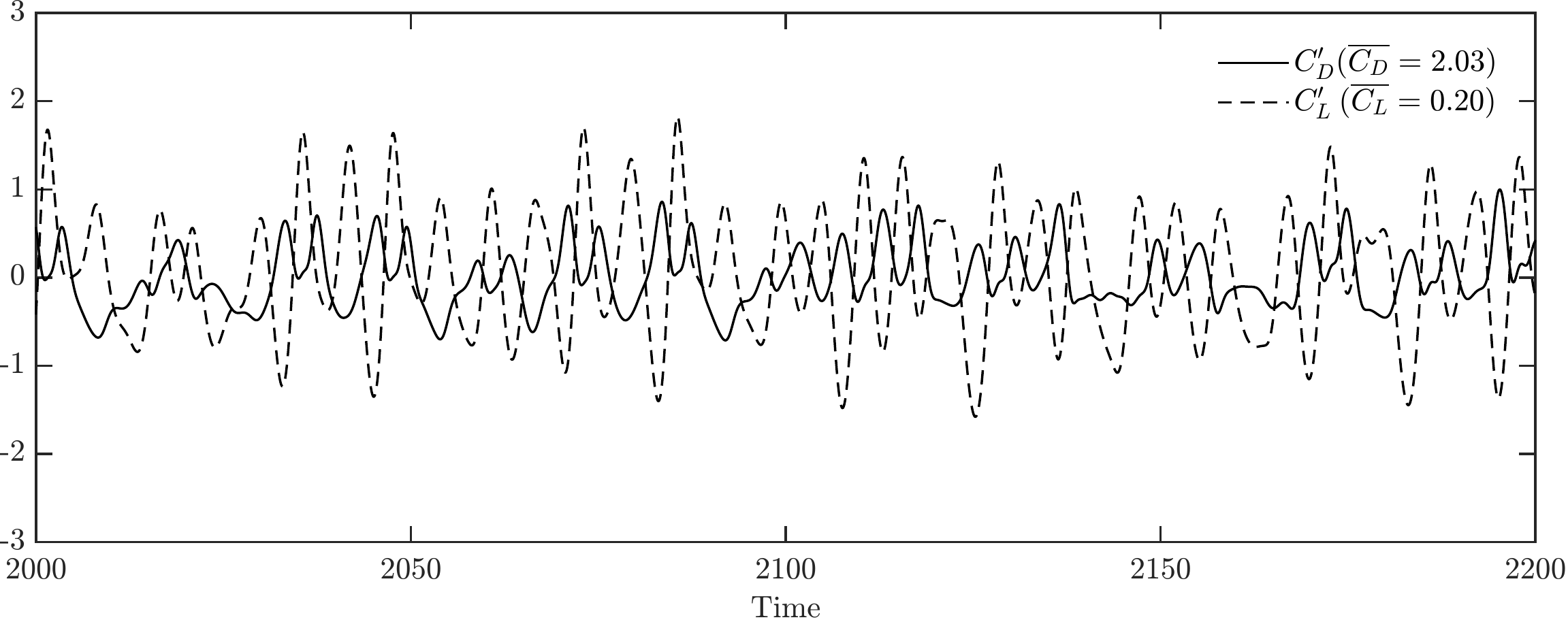}}
  \caption{Fluctuating lift (dashed line) and drag (solid line) coefficients of the top cylinder. Results are shown only from the last 200 time units.}
\label{fig:aero time history}
\end{figure}

\begin{figure}[ht]
  \centerline{\includegraphics[width=\textwidth]{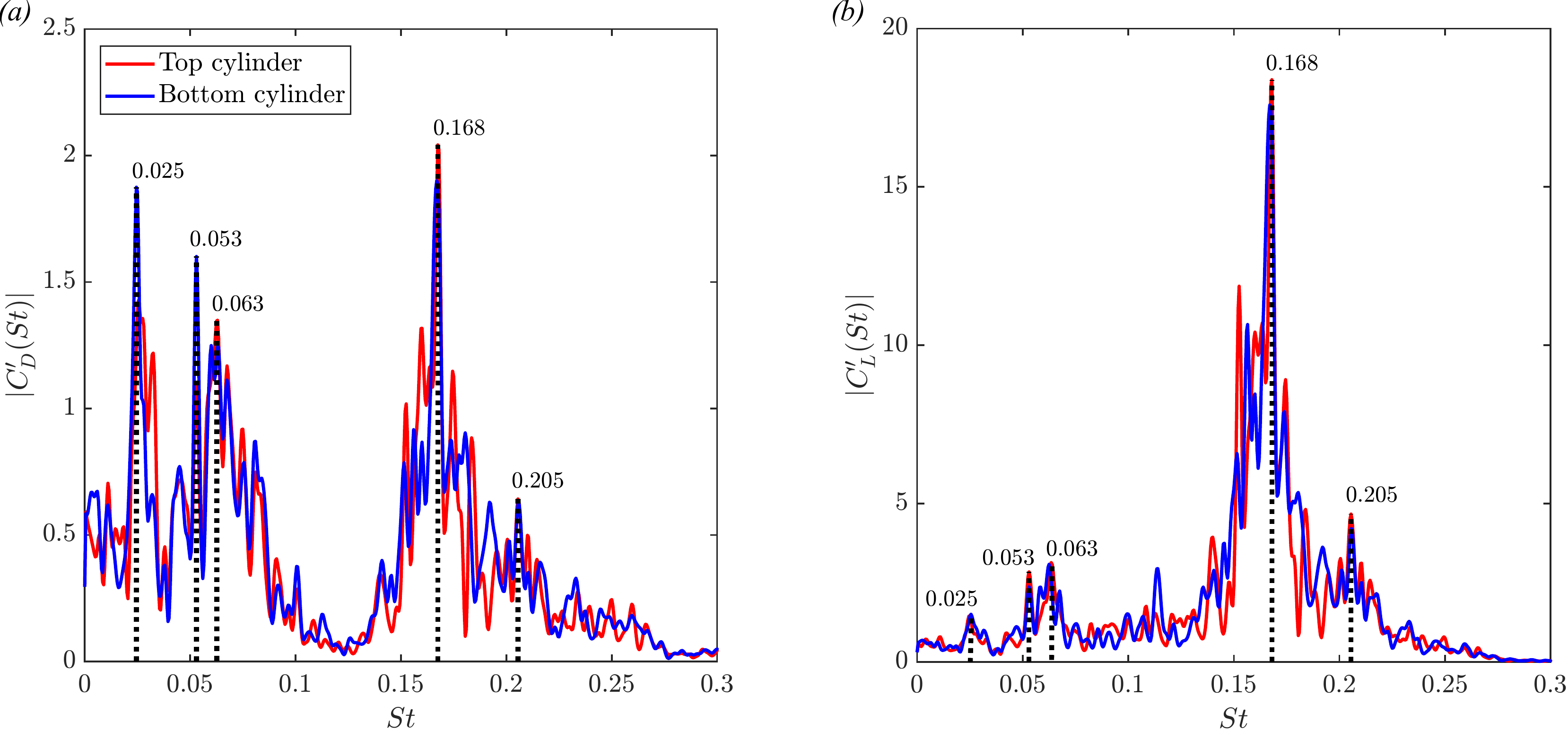}}
  \caption{PSD of the fluctuating drag ($a$) and lift ($b$) coefficients for both cylinders.}
\label{fig:aero spectra}
\end{figure}

The signals of $C'_L(t)$ and $C'_D(t)$ for the top cylinder are shown in figure \ref{fig:aero time history}. Both signals display a chaotic behaviour but with some underlying irregular periodicity. The  $C'_L(t)$ fluctuations are larger compared to those of $C'_D(t)$. The Power Spectral Densities (PSD) of the aerodynamic coefficients for both cylinders are plotted against the Strouhal number, $St=fD/U_{\infty}$, in figure \ref{fig:aero spectra}. Note the good matching for the top and bottom cylinders. While both spectra display a sharp peak at $St=0.168$, the spectrum of the drag coefficient is much richer, with pronounced peaks also at significantly lower frequencies, $St=0.025,\:0.053,\:0.063$. There is also a weaker peak at the higher frequency of $St=0.205$ for both coefficients. 

The two energetic frequency ranges identified by our simulations for the drag coefficient are in agreement with the results of \citet{Ma}. The numerical studies by \citet{Gera,Ma} identified the dominant shedding frequency for a single square cylinder to be $St \approx 0.15$ for $Re=200$. The present results show that a significant portion of the energy of the fluctuations is contained within the nearby frequency of $St=0.168$. The interaction between the two cylinders has resulted also in a multitude of slower temporal scales. Note that the  integration time of 2,200 time units, is long enough to resolve more than 50 cycles of the slowest temporal scale (with frequency $St=0.025$).

The spectra indicate a complex dynamical system with interacting multi-scale dynamics. The challenge of flow reconstruction is to deduce and predict this multitude of closely interacting temporal scales from sparse measurements. 

\begin{figure}[t]
  \centerline{\includegraphics[width=\textwidth]{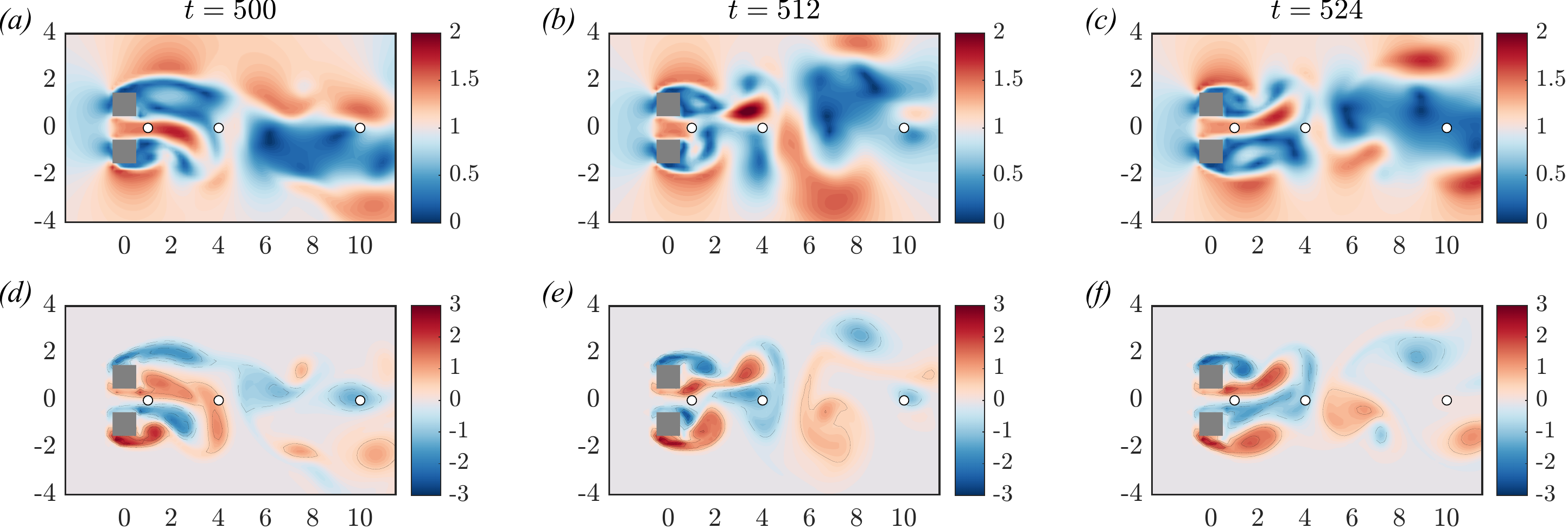}}
  \caption{Snapshots of velocity magnitude ($a$-$c$) and vorticity ($d$-$f$) fields at three time instants. The solid lines ($-$) indicate iso-contours of positive vorticity, while the dashed lines ($--$) iso-contours of negative vorticity. White circles ($\bigcirc$) indicate the three probe points (located at $x=1,4,10$) where velocity fluctuations are recorded; the corresponding spectra are shown in figure \ref{fig:velocity signals spectra}.}
\label{fig:snapshots}
\end{figure}

The irregular flapping motion of the jet emanating from the gap between the two cylinders plays a critical role in the flow development. This motion, which is analysed in detail in \citet{Ma}, results in an unsteady forcing on the vortical wakes developing behind the two cylinders; the mechanism is visualised in figure \ref{fig:snapshots} at 3 time instants. The contour plots of the instantaneous velocity magnitude (top row) demonstrate that the recirculation regions behind the cylinders (regions of low velocity magnitude are  coloured blue) are distorted by the gap flow (shown as an elongated patch of large velocity magnitude coloured red) and the phase determines the type of distortion. For example, at $t=500$ (left column) the gap flow is pointing downwards, thus the upper wake widens and the lower one shrinks. In turn, the asymmetric pressure field (resulting from the different strength of vorticity emanating from the top and bottom walls of the gap) is responsible for the flapping motion of the jet; see  vorticity contours at the bottom row of figure \ref{fig:snapshots}. For example at $t=500$, the counter-clockwise vortex from the top wall of the gap (shown in red) is leading the clockwise vortex from the bottom wall and the jet pitches down. At $t=524$ (right column), i.e.\ after 24 time units later (equal to roughly 4 periods of the main shedding cycle with $St=0.168$), the jet pitches up, and the sizes of the recirculation bubbles swap. In the intermediate time instant, $t=512$ (middle column), both recirculation regions appear similar in size and the jet has detached from the gap (see the small patch of large velocity at $x\approx 3-4$). It should be emphasised that this unsteady flapping motion is non-periodic, but over long time the two pitching directions of the jet have equal probability to occur, resulting in an overall symmetric flow pattern, as will be shown later.

\begin{figure}
  \centerline{\includegraphics[width=\textwidth]{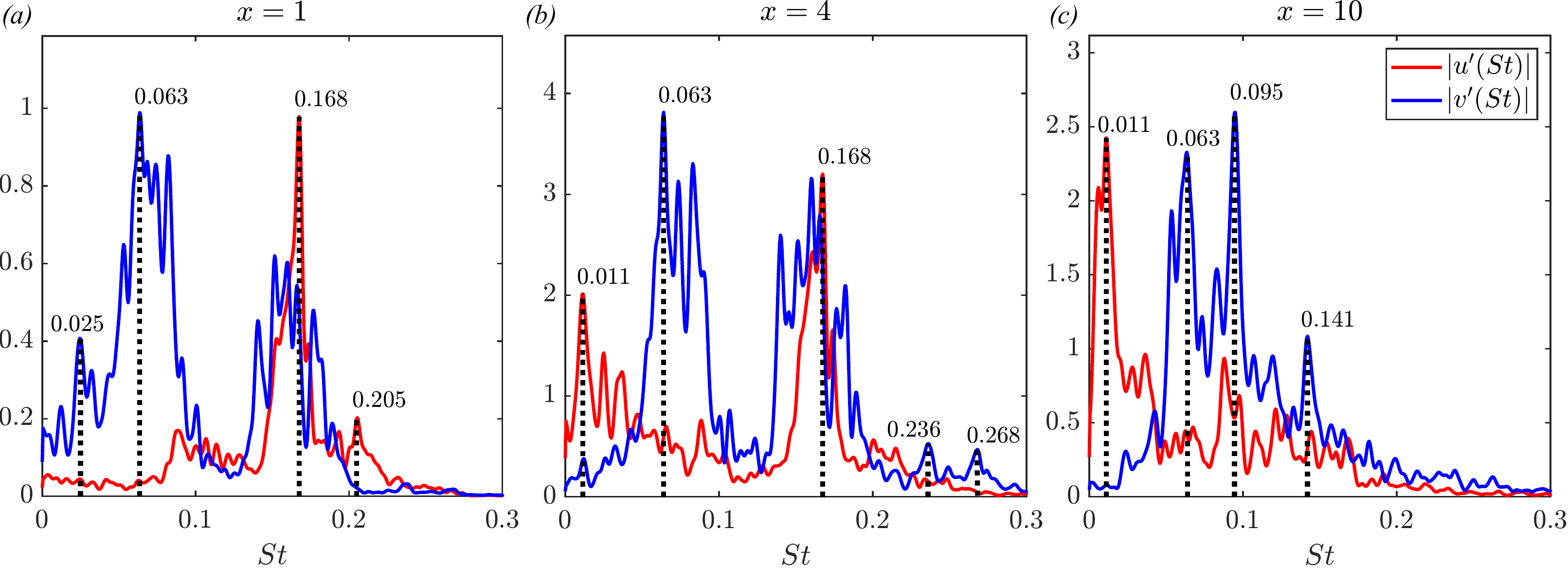}}
  \caption{PSD of fluctuating velocity components $u'$ (red) and $v'$ (blue) at three probe points on the centre line $y=0$ and at streamwise locations $x=1$ ($a$), $x=4$ ($b$) and $x=10$ ($c$).}
\label{fig:velocity signals spectra}
\end{figure}

In order to further investigate the irregular dynamics of the gap flow and the frequency content of the fully mixed downstream wake region, velocity fluctuations are extracted at three probe points placed along the centre line at locations $x=[1,4,10]$ (see white circle markers in figure \ref{fig:snapshots}). The resulting spectra are shown in figure \ref{fig:velocity signals spectra} and reveal some very interesting features. For the probe point located at $x=1$, i.e.\ very close to the gap between the cylinders, the $u'$ component has a single peak at $St=0.168$, but the $v'$ component peaks at a lower frequency, $St=0.063$ (but it also has significant energy content at $St=0.168$). This suggests that while the dynamics of vortex shedding leaves its signature on both components, the irregular dynamics of the flapping gap flow leaves its imprint very clearly only in the $v'$ component. Note also that the gap flow has energy content in a band of frequencies, from  $St \approx 0.05-0.09$. At $x=4$, the gap flow starts to mix with the two wakes forming at the rear of the cylinders. The $v'$ spectrum is similar to the one at $x=1$, but now the peak around $St=0.168$ becomes more pronounced, and the energy content is almost equally partitioned between two frequency bands, centred around $St=0.063$ and $St=0.168$.  On the other hand, the $u'$ spectrum retains the sharp peak at $St=0.168$ but now some small frequencies are starting to emerge, the most prominent one being $St=0.011$. This is probably a sub-harmonic of the $St=0.025$ appearing in $v'$ at $x=1$. At $x=10$, it is interesting to notice that the shedding frequency of $St=0.168$ is completely suppressed in both spectra. For the $u'$ component the small frequency of $St=0.011$, that has started to emerge upstream of $x=4$, now becomes dominant. The $v'$ component contains a mixture of low frequencies (that includes $St=0.063$), but some other new frequencies have also appeared, such as $St=0.095$ and $St=0.141$. The new frequencies arise from complex nonlinear interaction of Fourier components, for example $0.095\approx 0.063+3\times0.011$.  

The above analysis has clearly demonstrated that the mixing of the jet emerging from the gap with the vortical wakes significantly alters the velocity spectra. In the near wake, the flow is dominated by two characteristic frequencies, but as the flow evolves downstream the amalgamated wakes contain only low and intermediate frequencies. As explained in the introduction, most of the literature in flow reconstruction has focused on flows that are homogeneous in the streamwise direction, such as channel flow, that do not exhibit any of the flow features that dominate the present flow. 

\begin{figure}
  \centerline{\includegraphics[width=\textwidth]{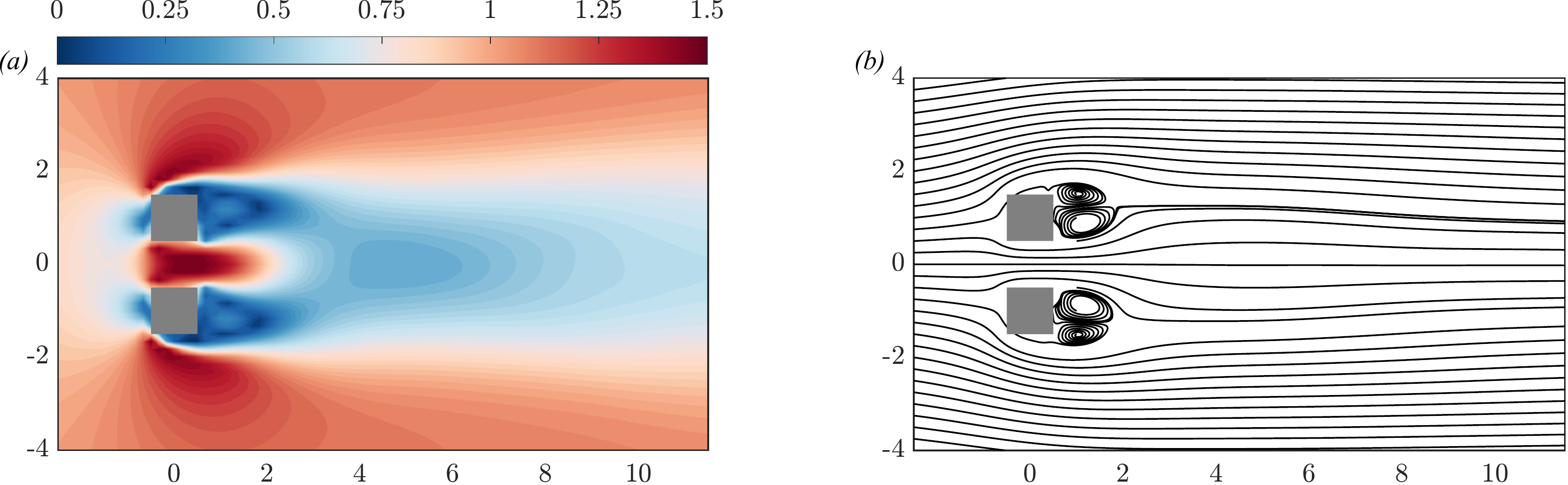}}
  \caption{Contours of velocity magnitude ($a$) and streamlines ($b$) of the time-averaged flow.}
\label{fig:time averaged flow}
\end{figure}

Although the flow exhibits strongly unsteady behaviour, the time-average flow pattern is symmetric. This is demonstrated in figure \ref{fig:time averaged flow} that shows contours of the velocity magnitude and streamlines of the mean flow. 

In the following section, we derive a reduced-order model of the flow. 

\section{Reduced-order model using POD}\label{sec:POD}

To obtain the POD modes, we first construct the snapshot matrix $\mathsfbi{Y}(\boldsymbol{x},t_1:t_K)$ by stacking the fluctuations $u'$ and $v'$ at different time instants column by column,
\begin{equation}
\setlength{\arraycolsep}{5pt}
\renewcommand{\arraystretch}{1.3}
\mathsfbi{Y}(\boldsymbol{x},t_1:t_K) = \left[
\begin{array}{cccc}
  u'^{(1)}(\boldsymbol{x}_{1})  &  u'^{(2)}(\boldsymbol{x}_{1}) & \cdots & u'^{(K)}(\boldsymbol{x}_{1}) \\
  \displaystyle
  \vdots & \vdots & \cdots & \vdots \\
  u'^{(1)}(\boldsymbol{x}_{\ell}) & u'^{(2)}(\boldsymbol{x}_{\ell}) & \cdots & u'^{(K)}(\boldsymbol{x}_{\ell}) \\ 
  \displaystyle
  v'^{(1)}(\boldsymbol{x}_{1}) & v'^{(2)}(\boldsymbol{x}_{1}) & \cdots & v'^{(K)}(\boldsymbol{x}_{1}) \\
  \displaystyle
  \vdots  & \vdots & \ddots  & \vdots \\
  v'^{(1)}(\boldsymbol{x}_{\ell}) & v'^{(2)}(\boldsymbol{x}_{\ell}) & \cdots & v'^{(K)}(\boldsymbol{x}_{\ell}) \\
\end{array}  \right],
\label{eq: snapshot matrix}
\end{equation}
where $\boldsymbol{x}_i=[x_i,\:y_i]$ is the position vector of the $i$-th spatial location where velocities are stored, $\ell$ denotes the total number of such locations, and $K$ is the number of snapshots. The points $\boldsymbol{x}_i \: (i=1\dots \ell)$ are within the area bounded by the red-dashed line shown in figure \ref{fig:geometry}. The velocities were interpolated on a uniform mesh with resolution $\Delta x = \Delta y = D/6$, resulting in a dataset with $\ell=4,067$.  The time window was 2000 time units and the step between two successive snapshots was $\Delta t = 0.04$, thus $K = 50,000$. The spectral analysis of the previous section showed that the energy content of the velocity fluctuations has $St\leq 0.30$. The aforementioned settings mean that a frequency $St=0.3$ is resolved with more than 80 snapshots, while the length of the signal covers 22 periods of the slowest temporal events with frequency $St=0.011$. 

The POD modes can be computed from singular value decomposition of the matrix $\mathcal{V}^{1/2}\mathsfbi{Y}$,  where $\mathcal{V} = (\Delta x \Delta y) \: \mathcal{I}_{2\ell \times 2\ell}$  and $\mathcal{I}_{2\ell \times 2\ell}$ is the identity matrix of size $2\ell \times 2\ell$; see \citet{H} for details about the underlying inner  product. Since $2\ell<K$, the singular value decomposition is written as  
\begin{equation}
\mathcal{V}^{1/2}\mathsfbi{Y} = \boldsymbol{\Phi}\boldsymbol{\Sigma}\boldsymbol{\Psi}^{\top},
    \label{eq: svd}
\end{equation}
\noindent where $\boldsymbol{\Phi} \in \mathbb{R}^{2\ell\times 2\ell}$ is the matrix of left singular vectors (containing the unscaled POD modes stacked into $K$ columns), $\boldsymbol{\Sigma}=diag(\sigma_{i}) \in \mathbb{R}^{2\ell\times 2\ell}$ is the diagonal matrix of singular values ranked in descending order and $\boldsymbol{\Psi} \in \mathbb{R}^{2\ell\times K}$ is the matrix of right singular vectors (containing the temporal dynamics of the modes). The economy-size \texttt{'econ'} option of \texttt{MATLAB} is used to compute only a small subset $m \ll 2\ell$ of the singular values. The scaled orthonormal spatial modes $\boldsymbol{\phi}_i(\boldsymbol{x})$ are the columns of 
\begin{equation}
    \boldsymbol{\phi}(\boldsymbol{x}) = \mathcal{V}^{-1/2} \boldsymbol{\Phi}.
    \label{eq: POD spatial modes}
\end{equation}
The energy content of each mode (eigenvalues) can be obtained from the singular values $\sigma_i$ from,
\begin{equation}
    \lambda_{i} = \frac{\sigma_{i}^2}{K}.
    \label{eq: eigvals}
\end{equation}
An approximation of the velocity perturbations can be computed from 
\begin{equation}
\left[
\begin{array}{c}
{u}' \\
{v}'\\
\end{array}
\right]
\left ( \boldsymbol{x},t \right )
\approx \sum_{i=1}^{m}a_i\left ( t \right )
\left[
\begin{array}{c}
{\phi}_i^{(u')} \\
{\phi}_i^{(v')}\\
\end{array}
\right]
\left ( \boldsymbol{x} \right ).
\label{eq: truncated_reconstruction}
\end{equation}
The temporal coefficients, $\boldsymbol{a}(t)=\left[a_1(t), a_2(t), \cdots, a_m(t) \right]$, are obtained from the projection operation,
\begin{equation}
\boldsymbol{a}(t) = \mathsfbi{Y}^{\top} \: \mathcal{V}^{1/2} \: \boldsymbol{\Phi}. 
    \label{eq: POD coefficients}
\end{equation}

\begin{figure}[ht]
  \centerline{\includegraphics[width=\textwidth]{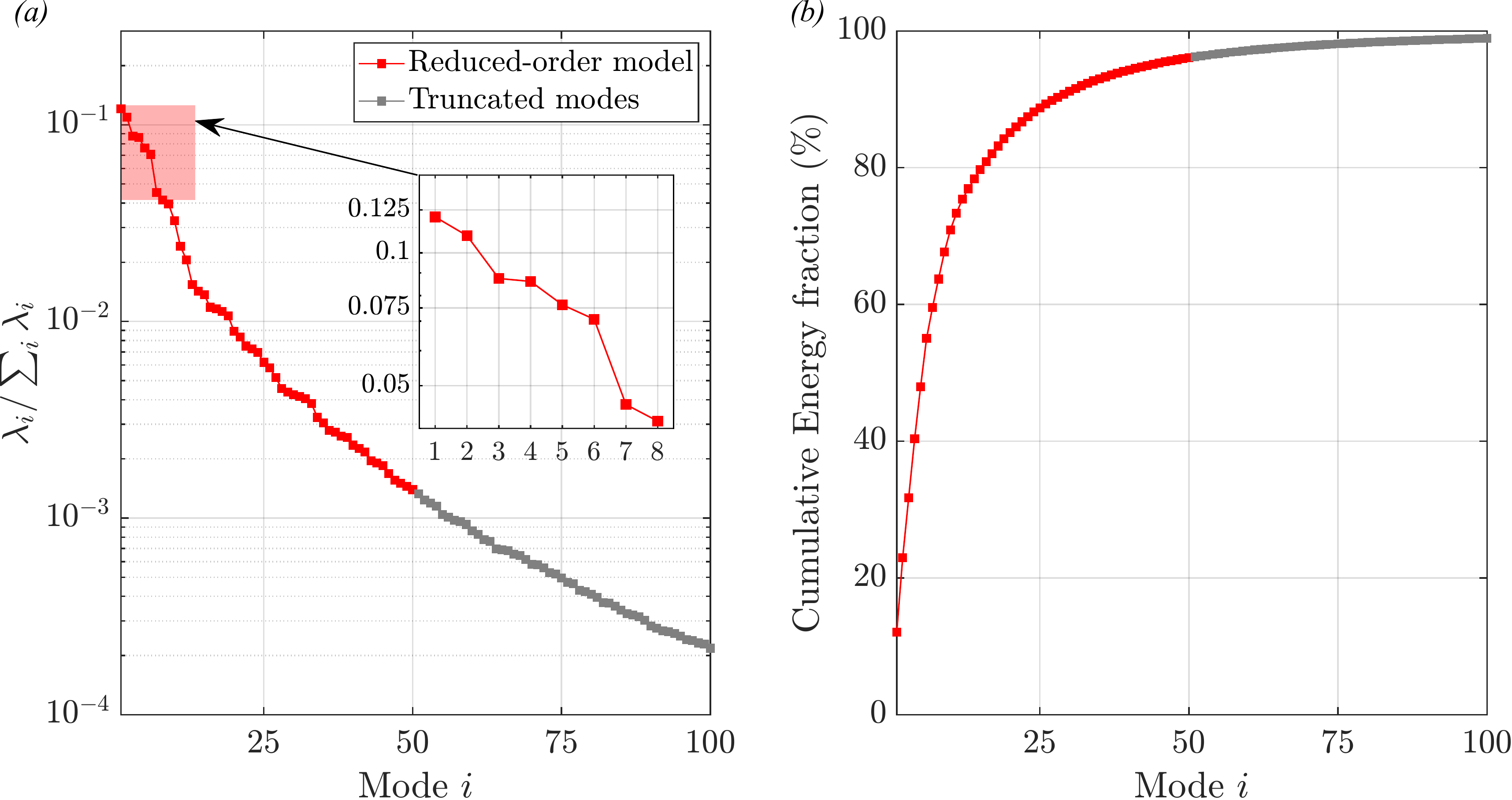}}
  \caption{Distribution of the energy content of the first 100 eigenvalues ($a$) and their cumulative sum ($b$). The inset in ($a$) shows the leading 8 eigenvalues. Modes retained for construction of the reduced-order model are coloured in red, while truncated modes in grey.}
\label{fig:eigvals}
\end{figure}

The normalised eigenvalues $\lambda_i$ and their cumulative sum are plotted in figure \ref{fig:eigvals}. In contrast to flows with simpler dynamics, such as those studied in \citet{Ooi} and \citet{Inigo_Sodar_Papadakis_2019}, the energy spectrum of the examined flow decays more slowly. In total, 50 modes are required to capture approximately 97\% of the kinetic energy. Under regular vortex shedding conditions, the POD modes appear in pairs, with the two members of each pair having exactly the same energy, see for example \citet{Inigo_Sodar_Papadakis_2019}. In the present flow case, there is weak mode pairing between modes 1\&2, 3\&4, 5\&6 and 7\&8 as can be seen from the left panel of figure \ref{fig:eigvals} (and especially the inset). The energies between two paired modes are close, but not exactly the same, probably due to the irregular nature of the shedding pattern.  

\begin{figure}
  \centerline{\includegraphics[width=\textwidth]{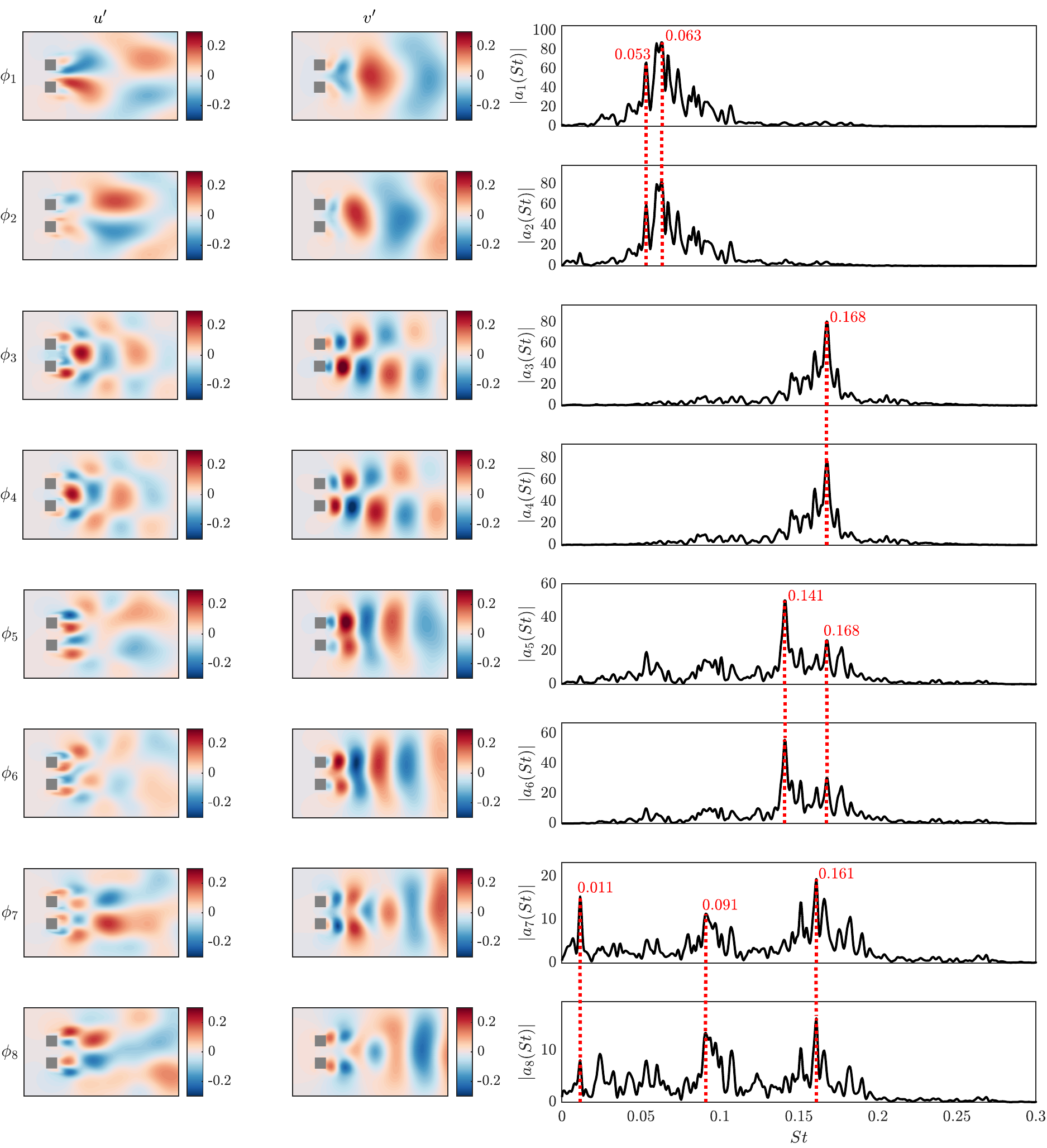}}
  \caption{Leading 8 POD modes, containing $\sim 64\%$ of the total fluctuating energy. The spatial distributions of the $u'$ and $v'$ eigenmodes are shown on the left and middle columns respectively. The PSD spectra of the associated time coefficients are shown on the right column. The dominant frequencies are marked with vertical, red-dashed lines that extend over a pair of modes.}
\label{fig:modes}
\end{figure}

The mode shapes $\phi_i^{(u')}(\boldsymbol{x}), \phi_i^{(v')}(\boldsymbol{x})$ and the frequency content of the temporal coefficients $a_i(t)$ of the leading 8 modes are shown in figure \ref{fig:modes} (left, middle and right columns respectively). The first two modes capture the low-frequency range of the spectrum ($St=0.053,0.063$), which characterises the flapping gap flow (see spectrum of $v'$ at $x=1$ and $x=4$ in figure \ref{fig:velocity signals spectra}). The  $\phi_1^{(u')}(\boldsymbol{x})$ field shows large anti-symmetric patches that originate at the gap, while the $\phi_1^{(v')}(\boldsymbol{x})$ structure is maximised at the centre line, indicating  symmetry breaking. There is spatial shift between the first two modes, most easily recognised by comparing the $v'$ fields, i.e.\ $\phi_1^{(v')}(\boldsymbol{x})$ and  $\phi_2^{(v')}(\boldsymbol{x})$.

The spectra of the second mode pair (modes 3-4) peak at $St=0.168$, which is the frequency associated with the main vortex shedding cycle. This can be ascertained by inspecting mainly the $\phi_3^{(v')}(\boldsymbol{x})$ and $\phi_4^{(v')}(\boldsymbol{x})$ fields; each contains two branches of $v'$ structures that peak behind the cylinders, indicating loss of symmetry for either wake. The structures grow downstream and their centres move away from the centre-line. The picture corresponds to out-of-phase shedding activity behind the two cylinders. Note again the spatial shift between the two fields; this shift combined with a sharp frequency peak indicates  propagating structures, typical of shedding activity.  The POD decomposition has thus succeeded in capturing the two most important mechanisms using only four modes. These modes contain approximately $40\%$ of the kinetic energy of the fluctuations (see cumulative energy plot in figure \ref{fig:eigvals}). 

The spectra of the third mode pair (modes 5-6) peak at  $St=0.141$, a frequency that was also observed in the $v'$ spectra at $x=10$. The vertically aligned structures with the same sign in the $\phi_5^{(v')}(\boldsymbol{x})$ and $\phi_6^{(v')}(\boldsymbol{x})$ fields indicate in-phase shedding. These structures are initially distinct behind each cylinder, but further downstream they amalgamate and form a single diffusing structure.  Modes 7 and 8 have broader spectrum and less obvious spatial structure, thus become more difficult to interpret. Slow dynamics start to appear, as can be seen by the emergence of a peak at $St=0.011$. Other temporal scales in the moderate frequency range, $St=0.091,0.161$, also emerge. This suggests that these modes capture structures arising from nonlinear interactions in the wake.  For higher modes (results not shown), the spectra become even more noisy and interpretability is lost.

The objective now is to construct a data-driven, linear dynamic estimator for the fluctuations $u'\left ( \boldsymbol{x},t \right )$ and $v'\left ( \boldsymbol{x},t \right )$ from the velocity signals at a few sparsely located sensors (the question of sensor placement is considered in \S\ref{sec:Sparse sensor placement}). Since the fluctuations can be described accurately  by the POD expansion \eqref{eq: truncated_reconstruction}, we proceed with deriving an estimator for $a(t)$; this makes the problem tractable because of the low dimensionality of $a(t)$.  
\section{Construction of a data-driven estimator using system identification and optimal estimation theory\label{sec:Flow reconstruction using system identification and optimal estimation theory}}

We start by deriving the evolution equation for $a(t)$. To this end, we consider the governing equations of the fluctuations and keep the linear terms in the left-hand side,   
\begin{subequations}
\begin{equation}
  \underbrace{\frac{\partial \boldsymbol{u}'}{\partial t} + \boldsymbol{u}' \cdot \nabla \overline{ \boldsymbol{u}}  + \overline{\boldsymbol{u}} \cdot \nabla \boldsymbol{u}' + \nabla p' - Re^{-1} \: \nabla^2 \boldsymbol{u}'}_{\textup{Linear terms in} \:\: \boldsymbol{u}'} = \underbrace{\overline {\boldsymbol{u}' \cdot \nabla \boldsymbol{u}'}  - \boldsymbol{u}' \cdot \nabla \boldsymbol{u}'}_{\textup{Nonlinear terms in} \:\: \boldsymbol{u}'},
  \label{eq: evolution equation of fluctuations}
\end{equation}
\begin{equation}
  \nabla \cdot \boldsymbol{u}' = 0,
\end{equation} 
\label{eq: NS of fluctuations}
\end{subequations}
\noindent where $\boldsymbol{u}'=\left[ u' \:\: v' \right]^{\top}$. In the above equation set, we substitute the complete POD expansion of the fluctuation field, 
\begin{equation}
     \boldsymbol{u}'\left ( \boldsymbol{x},t \right )=  
\sum_{i=1}^{\infty}a_i\left ( t \right ) \boldsymbol{\phi}_i \left ( \boldsymbol{x} \right ) 
\quad \mbox{or} \quad
 \left[
\begin{array}{c}
{u}' \\
{v}'\\
\end{array}
\right]\left ( \boldsymbol{x},t \right )
=\sum_{i=1}^{\infty}a_i\left ( t \right )
\left[
\begin{array}{c}
{\phi}_i^{(u')} \\
{\phi}_i^{(v')}\\
\end{array}
\right]\left(\boldsymbol{x}\right),
     \label{eq:full_reconstruction}
\end{equation}
\noindent and then perform Galerkin projection. Exploiting the divergence-free and orthonormal properties of $\boldsymbol{\phi}_i(\boldsymbol{x})$, we obtain the evolution equation of the temporal coefficients of the dominant $m$ modes $\boldsymbol{a}(t)=\left[a_1(t), a_2(t), \cdots, a_m(t) \right]$, 
\begin{equation}
\frac{\textup{d}\boldsymbol{a}}{\textup{d}t} = \mathcal{A}\:\boldsymbol{a}(t) + \mathcal{F}(t)+\mathcal{\epsilon}(t),
    \label{eq: CT-evolution of a(t)}
\end{equation}
\noindent where matrix $\mathcal{A}$ originates from the linear terms of \eqref{eq: evolution equation of fluctuations}, the   forcing term $\mathcal{F}(t)$ comes from the nonlinear terms, and $\mathcal{\epsilon}(t)$ arises because the truncated temporal vector $\boldsymbol{a}(t)$ contains only $m$ elements, see \citet{Inigo_Sodar_Papadakis_2019}. In the following, we select the truncation order to be $m=50$ (this mode range is marked in red colour in  figure \ref{fig:eigvals}). Note that the $i$-th element of the vector $\mathcal{F}(t)$ takes the form 
\begin{equation}
    \mathcal{F}_i(t)= \sum_{j=1}^{m} \sum_{k=1}^{m} C_{ijk} a_j(t)a_k(t) 
 \label{eq: F-vector}   
\end{equation}

In discrete form, \eqref{eq: CT-evolution of a(t)} can be written as \begin{equation}
    \boldsymbol{a}[k+1] = \mathcal{A}'\: \boldsymbol{a}[k] + \mathcal{F}'[k]+\mathcal{\epsilon}'[k],
 \label{eq: DT-evolution of a(t)}     
\end{equation}

We now apply system identification and seek a linear model of the form 
\begin{equation}
    \tilde{\boldsymbol{a}}[k+1] = \mathsfbi{A} \tilde{\boldsymbol{a}}[k] + \tilde{\boldsymbol{f}}[k],
 \label{eq: identified-evolution of a(t)}   
\end{equation}
where $\tilde{\boldsymbol{f}}[k]$ is a noise term, that will approximate \eqref{eq: CT-evolution of a(t)}. The model can be more generally written as 
\begin{subequations}
\begin{equation}
    \boldsymbol{x}[k+1] = \mathsfbi{A} \boldsymbol{x}[k] + \boldsymbol{w}[k],
    \label{eq: plant_a}
\end{equation}
\begin{equation}
    {\boldsymbol{a}}[k] = \mathsfbi{C} \boldsymbol{x}[k] + \boldsymbol{v}[k],
\end{equation}
\label{eq: plant}
\end{subequations}
\noindent where we have introduced the internal state vector $\boldsymbol{x}[k]$, which does not necessarily have the same dimension as the output $\boldsymbol{a}[k]$. This is known as the \textit{process} state-space form of a discrete, linear, time-invariant system with process noise $\boldsymbol{w}[k]$,  output noise $\boldsymbol{v}[k]$, and no input.

Before proceeding further, it is very instructive to reflect on the form of the model we seek \eqref{eq: identified-evolution of a(t)} and the true model \eqref{eq: DT-evolution of a(t)}. If the perturbations $\boldsymbol{u}'\left ( \boldsymbol{x},t \right )$ were infinitesimal, as in \cite{guzman2014dynamic}, then $\mathcal{F}_i(t)=0$ and both models are linear. However, if the perturbations are finite, as in the present case, then $\mathcal{F}_i(t) \ne 0$, and \eqref{eq: identified-evolution of a(t)} requires careful physical interpretation. Here we argue that system identification approximates the true $\mathcal{F}_i(t)$ with the linear form $\mathcal{F}_i(t) \approx \sum_{j=1}^{m} B_{ij} a_j(t)$, so that  $\tilde{\boldsymbol{a}}[k]$ optimally matches with ${\boldsymbol{a}}[k]$. This is an eddy-viscosity type approximation in the domain of POD time coefficients, where the matrix elements $B_{ij}$ are obtained directly from data. This is an important advantage of data-driven methods, because the functional form of the eddy-viscosity is generally not known. Errors arising from this approximation are included in the noise term, $\tilde{\boldsymbol{f}}[k]$.

We use the system identification algorithm  \texttt{n4sid}, to identify the pair $\left \{\mathsfbi{A},\: \mathsfbi{C}\right \}$ in \eqref{eq: plant} from the true output signal, $\boldsymbol{a}[k]$, obtained from equation \eqref{eq: POD coefficients}; refer to \cite{Overscheereport}, the books of \cite{O,ljung1999system} and the review paper of \cite {Qin} for more details about this algorithm. The SLICOT package, see  \cite{Sima_et_al_2004} and references therein, can also provide the covariance matrices of the noise sequences $\boldsymbol{w}[k]$ and $\boldsymbol{v}[k]$ (and also the cross-covariance matrix). The model order, $n$, of system \eqref{eq: plant} is also unknown and can be either computed as part of the solution or pre-specified. Note that $n$ is different from the number of retained POD modes $m$; more details about the relation of $n$ and $m$ will be provided in section \ref{sec:Performance assessment of dynamic estimation}. 

In the present work, we used the \texttt{n4sid} command of \texttt{MATLAB}, that formulates the plant in the following \textit{innovation} form:
\begin{subequations}
\begin{equation}
    \boldsymbol{x}[k+1] = \mathsfbi{A}\boldsymbol{x}[k] + \mathsfbi{K}\boldsymbol{e}[k],
\end{equation}
\begin{equation}
    \boldsymbol{a}[k] = \mathsfbi{C}\boldsymbol{x}[k] + \boldsymbol{e}[k],
\end{equation}
\label{eq: plant innovation form}
\end{subequations}
\noindent i.e.\ the process noise $\boldsymbol{w}[k]$ is related to the innovation vector $\boldsymbol{e}[k]=\boldsymbol{a}[k] - \mathsfbi{C}\boldsymbol{x}[k]$ with
\begin{equation}
\boldsymbol{w}[k] = \mathsfbi{K}\boldsymbol{e}[k],
    \label{eq: innovation vector vs process noise}
\end{equation}
\noindent where $\mathsfbi{K}$ is a gain matrix; see \cite{Qin} for the steps that lead from \eqref{eq: plant} to \eqref{eq: plant innovation form}. The covariance of the process noise  $\mathsfbi{Q} =\textup{cov} \left (\boldsymbol{w} \right ) = \mathbb{E}  \left ( \boldsymbol{w}\boldsymbol{w}^{\top} \right )$ can be computed from 
\begin{equation}
    \mathsfbi{Q} = \textup{cov} \left (  \mathsfbi{K}\boldsymbol{e} \right ) = \mathsfbi{K} \:  \textup{cov} \left (  \boldsymbol{e} \right ) \mathsfbi{K}^{\top}=\mathsfbi{K} \: \mathbb{E}  \left ( \boldsymbol{e}\boldsymbol{e}^{\top} \right )  \mathsfbi{K}^{\top},
    \label{eq:Q_covariance}
\end{equation}
where $\mathbb{E}$ is the expectation operator and $\mathbb{E} \left ( \boldsymbol{e}\boldsymbol{e}^{\top} \right )=\left(\sum_{k=1}^{k=K}\boldsymbol{e}[k]\boldsymbol{e}^\top[k]\right)/K$. 

It is now assumed that $u'$ or $v'$ velocities can be measured at $p$ sensors, and that each sensor can record only one velocity component. Thus, there are $2\ell$ potential sensors. The measurements are stored in vector $\boldsymbol{s}[k]$. For example, if all $p$ sensors measure  $u'$, then $\boldsymbol{s}[k]=\left [ u_1'[k], u_2'[k], \cdots ,u_p'[k] \right] ^{\top}$.  Vector  $\boldsymbol{s}[k]$ can be written in terms of the POD temporal coefficients as: 
\begin{equation}
\boldsymbol{s}[k] = \mathsfbi{S} \boldsymbol{a}[k] + \boldsymbol{g}[k],
    \label{eq: input with noise}
\end{equation}
\noindent where matrix $\mathsfbi{S}$ consists of the rows of $\boldsymbol{\phi}({\boldsymbol{x}})$ (see \eqref{eq: POD spatial modes}) corresponding to the locations $\boldsymbol{x}_j$ and velocity component(s) recorded at each location. In matrix form  $\mathsfbi{S}=\mathbb{S}_{\boldsymbol{\phi}}\boldsymbol{\phi}({\boldsymbol{x}})$, where $\mathbb{S}_{\boldsymbol{\phi}}$ 
is the row-selection matrix, that consists of $p$ rows and $2\ell$ columns.  At each row of $\mathbb{S}_{\boldsymbol{\phi}}$, all the elements are set to $0$, except for the element that corresponds to the index of the selected sensor, which takes the value of 1. For example, if there is just one sensor which is the third among $2\ell$ potential sensors, $\mathbb{S}_{\boldsymbol{\phi}}$ is the $1 \times 2l$ matrix $\mathbb{S}_{\boldsymbol{\phi}}=\left[0 \; 0 \; 1 \; 0 \dots 0 \right] $. If $\mathbb{S}_{\boldsymbol{\phi}}$ is the identity matrix, $\mathbb{S}_{\boldsymbol{\phi}}=\mathbb{I}_{2\ell \times 2\ell}$,  measurements of both velocity components are recorded at all $\ell$ points. The problem of sensor placement amounts to specifying matrix $\mathbb{S}_{\boldsymbol{\phi}}$ and will be analysed in more detail in section  \ref{sec:Sparse sensor placement} below.  Equation \eqref{eq: input with noise} can be written as 
\begin{equation}
\boldsymbol{s}[k] = \mathbb{S}_{\boldsymbol{\phi}}\boldsymbol{\phi}({\boldsymbol{x}}) \boldsymbol{a}[k] + \boldsymbol{g}[k],
    \label{eq: input with noise_2}
\end{equation}
The noise term $\boldsymbol{g}[k]$ stems from the POD truncation and can also include the sensor measurement noise. The noise covariance $\mathsfbi{R}$ can be also computed from the available dataset from,
\begin{equation}
\mathsfbi{R} = \mathbb{E}  \left ( \boldsymbol{g}\boldsymbol{g}^{\top} \right ), \mbox{ where } \boldsymbol{g}[k]= \boldsymbol{s}[k] - \mathsfbi{S}\boldsymbol{a}[k].
\label{eq: measurement noise covariance}
\end{equation}

We are now ready to formulate the Kalman filter estimator: 
\begin{subequations}
\begin{equation}
    \hat{\boldsymbol{x}}[k+1] = \mathsfbi{A} \hat{\boldsymbol{x}}[k] + \mathcal{L} \left (  \boldsymbol{s}[k] - \hat{\boldsymbol{s}}[k]\right ),
    \label{eq: kalman filter for x}
\end{equation}
\begin{equation}
\hat{\boldsymbol{a}}[k] = \mathsfbi{C} \hat{\boldsymbol{x}}[k],
 \label{eq: kalman filter for a}
\end{equation}
\begin{equation}
\hat{\boldsymbol{s}}[k]=\mathsfbi{S}\hat{\boldsymbol{a}}[k] = \mathsfbi{S}\mathsfbi{C} \hat{\boldsymbol{x}}[k],
 \label{eq: kalman filter for s}
\end{equation}
\label{eq: kalman filter formulation}
\end{subequations}
\noindent where $\boldsymbol{s}[k]$ and $\hat{\boldsymbol{s}}[k]$ are the true  and estimated measurements respectively. The steady-state Kalman filter gain $\mathcal{L}$ can be obtained from the  solution to an algebraic Riccati equation \citep{Anderson_and_moore}, and the \texttt{MATLAB} function \texttt{kalman} was used.

The whole process can be summarised as follows:
\begin{enumerate}

\item Perform DNS simulations, assemble the snapshot matrix $\mathsfbi{Y}(\boldsymbol{x},t_1:t_K)$ and obtain the mode matrix $\boldsymbol{\phi}(\boldsymbol{x})$ and temporal coefficients $\boldsymbol{a}(t)=\left[a_1(t), a_2(t), \cdots, a_m(t) \right]^{\top}$. 

\item Assemble the measurement vector $\boldsymbol{s}[k]$ from the velocity fluctuations $u'$ or $v'$ at $p$ sensors. 

\item Provide $\boldsymbol{a}[k]$ to the system identification algorithm \texttt{n4sid} and extract the matrices $ \mathsfbi{A}, \: \mathsfbi{C}$ (to within a similarity transformation) and $\mathsfbi{K}$. Obtain covariance matrices, $\mathsfbi{Q}$  from \eqref{eq:Q_covariance}, and $\mathsfbi{R}$ from  \eqref{eq: measurement noise covariance}. 

\item Using matrices $\mathsfbi{A}$, $\mathsfbi{C}$, $\mathsfbi{S}$, $\mathsfbi{Q}$, $\mathsfbi{R}$, solve the algebraic Riccati equation to obtain the filter gain  $\mathcal{L}$.   

\end{enumerate}

Step (ii) requires the strategic selection of the sensor locations and the velocity component(s) measured at each location; this will be considered in more detail in section \ref{sec:Sparse sensor placement}. The central part of step (iii) is the singular value decomposition of a block-Hankel matrix assembled from the output data, $\boldsymbol{a}[k]$. 

Steps (ii)-(iv) are performed using data from a training dataset (in the present case 25,000 snapshots,  i.e.\ half of total number, corresponding to 1,000 time units). Once the estimator has been obtained, its performance is assessed by applying it to a new set of data, the validation dataset (the remaining 25,000 snapshots). The error between $\boldsymbol{\hat{a}}[k]$ and $\boldsymbol{a}[k]$ is defined as the percentage FIT between the predicted $\hat{a}_i[k]$ and the true $a_{i}[k]$ coefficient of the  $i$-th mode, 
\begin{equation}
	\textup{FIT}_i\: [\%] = 100 \left (1 -  \frac{ \left \| a_i[k] -  \hat{a}_i[k] \right \|}{\left \| a_i[k] -   \overline{a_i[k]} \right \|} \right ).
	\label{eq: FIT of POD modes}
\end{equation}
This parameter is equal to 100\% for perfect prediction and can become negative in case of very poor prediction. In the above, $\left \| \cdot \right \|$ denotes the $\mathcal{L}_2$ norm of the time signals.

\begin{figure}[ht]
  \centerline{\includegraphics[width=0.95\textwidth]{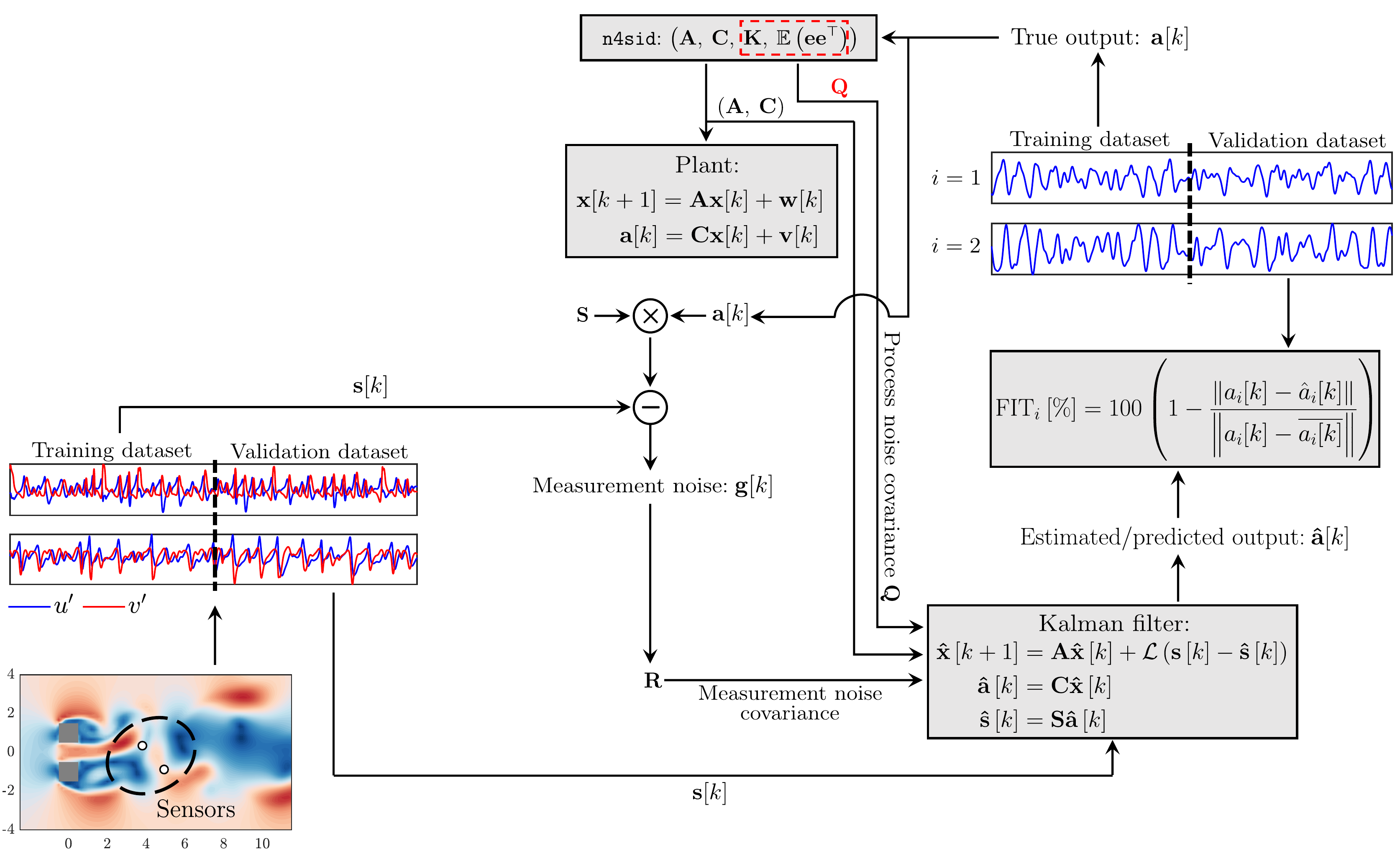}}
  \caption{Schematic of the estimator using stochastic system identification and Kalman filter.}
\label{fig:kalman filter}
\end{figure}

In the procedure  described above, the underlying system is first identified, and then a optimal estimator is constructed using Kalman filter. A schematic representation of this procedure is shown in figure \ref{fig:kalman filter}. It is also possible to extract the estimator directly from the input/output data, as in \cite{Inigo_Sodar_Papadakis_2019,Mikhaylov_et_al_2021}; this approach is also briefly described below. 

\section{Direct identification of the dynamic estimator\label{sec:Direct identification of the dynamical estimator}}
\begin{figure}[ht]
  \centerline{\includegraphics[width=\textwidth]{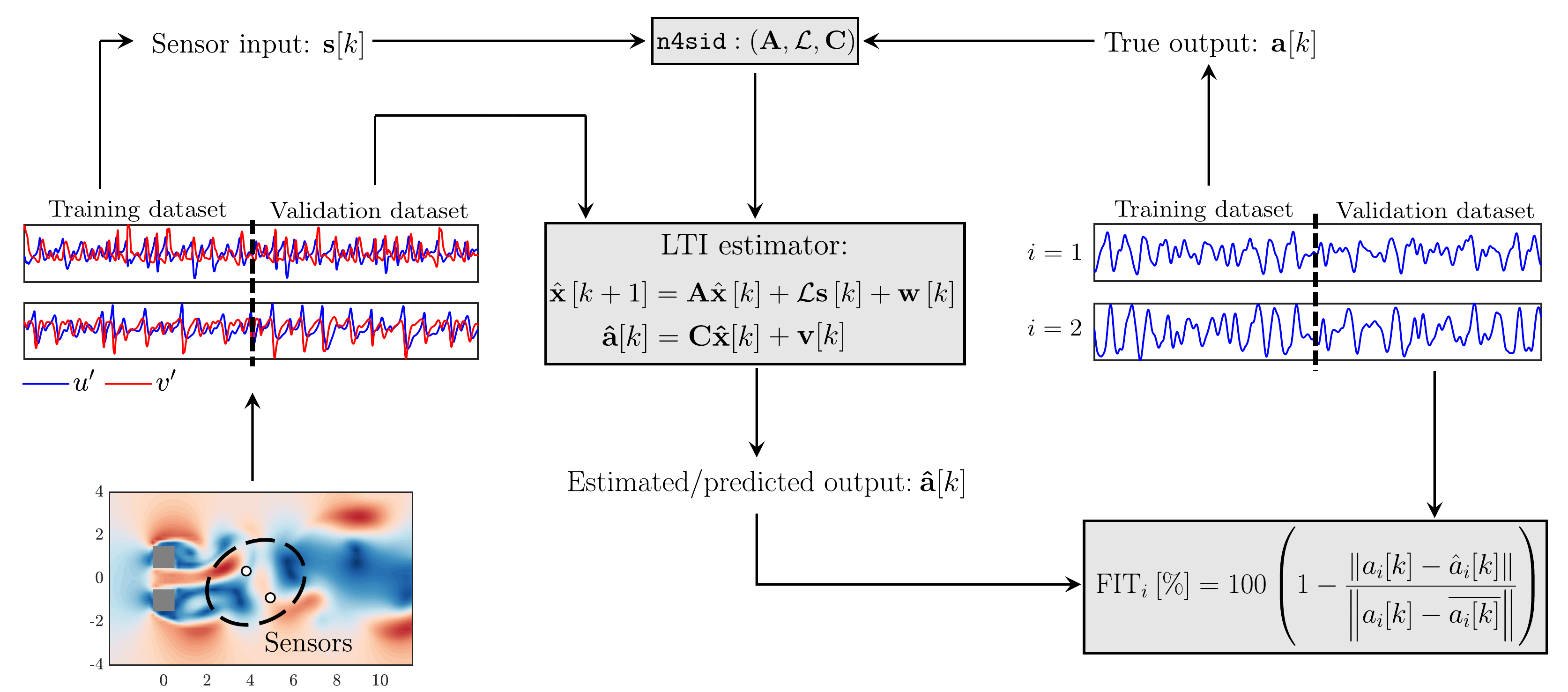}}
  \caption{Schematic of the direct identification approach for the dynamic estimator.}
\label{fig:original estimator}
\end{figure}

Substituting \eqref{eq: kalman filter for s} to \eqref{eq: kalman filter for x} we get:
\begin{subequations}
\begin{equation}
\hat{\boldsymbol{x}}[k+1] =\underbrace{\left( \mathsfbi{A}- \mathcal{L} \mathsfbi{S}\mathsfbi{C} \right)}_{=\mathcal{A}} \hat{\boldsymbol{x}}[k] + \mathcal{L} {\boldsymbol{s}}[k],
\end{equation}
\begin{equation}
\boldsymbol{\hat{a}}[k] = \mathsfbi{C} \boldsymbol{\hat{x}}[k],
\end{equation}
\label{eq: simple estimator}
\end{subequations}
or in more general form:
\begin{subequations}
\begin{equation}
\hat{\boldsymbol{x}}[k+1] = \mathcal{A} \hat{\boldsymbol{x}}[k] + \mathcal{L} {\boldsymbol{s}}[k]+\boldsymbol{w}[k],
\end{equation}
\begin{equation}
\boldsymbol{\hat{a}}[k] = \mathsfbi{C} \boldsymbol{\hat{x}}[k]+\boldsymbol{v}[k],
\end{equation}
\label{eq: simple estimator with noise}
\end{subequations}
\noindent where we have added stochastic white noise terms $\boldsymbol{w}[k]$ and $\boldsymbol{v}[k]$ in the system dynamics and the output. Again we use the \texttt{n4sid} command of \texttt{MATLAB} Toolbox \citep{ljung1999system} to identify the matrices $\mathcal{A}$, $\mathcal{L}$ and $\mathsfbi{C}$ from the input data $\boldsymbol{s}[k]$ and the output data $\boldsymbol{a}[k]$. A schematic overview of direct identification is shown in figure \ref{fig:original estimator}. 

Identifying directly the matrices of the estimator \eqref{eq: simple estimator with noise} takes significantly more time compared to identifying the underlying dynamical system \eqref{eq: plant innovation form}. This is because now there are two block-Hankel matrices, one for the input data, and one for the output data. For example, the runtime for direct identification with $n=50$ states and $p=50$  sensors was in excess of 4 days on an Intel(R) Xeon(R) CPU E5-2698 v4 @ 2.20GHz processor (serial computations), while the two-step approach described in section \ref{sec:Flow reconstruction using system identification and optimal estimation theory} took approximately $15$ minutes in the same computer.

The two-step construction approach extracts a lot more information from the training dataset compared to the direct approach. For example, it (approximately) accounts for the statistics of the forcing term $\mathcal{F}(t)$ in \ref{eq: CT-evolution of a(t)} (through the covariance of the process noise $\mathsfbi{Q}$), the effect of POD truncation (through the covariance of the measurement noise $\mathsfbi{R}$) and the shape of the POD modes (through matrix $\mathsfbi{S}$). The performance of the two approaches is compared in section \ref{sec:Performance assessment of dynamic estimation}. We first however consider the placement of the sparse sensors; this is the subject of the next section.


\section{Sparse sensor placement\label{sec:Sparse sensor placement}}

The parameters that are expected to affect the reconstruction quality are the model order, $n$, and the number, $p$, and location of sensors. As shown in \citet{Loiseau}, \citet{Manohar2} and \citet{Bhattacharjee}, a group of $p$ strategically located sensors can substantially improve the quality of estimation of complex systems.  Optimal sensor placement is an NP-hard problem \citep{Bnew}, i.e.\ there are $ \left(\begin{matrix} N \\ P \end{matrix} \right )= \frac{N!}{(N-P)!}$ combinations of placing P sensors at N potential locations (in our case $\textup{P}=p$, and $\textup{N}=2\ell$). For each combination, an estimator is constructed and its performance accessed. For very small number of sensors, say $p=1$ or $2$, this optimisation problem can be solved, see \citet{Inigo_Sodar_Papadakis_2019}, but very quickly becomes intractable.  Thus, an approximate solution is required.

In this paper, we adapt the greedy QR pivoting algorithm of  \citet{Manohar_et_al_2021} to identify sub-optimal sensors. Consider the fully-sensed (i.e.\ $\mathbb{S}_{\boldsymbol{\phi}}=\mathbb{I}_{2\ell\times 2\ell}$) discrete time system, $\boldsymbol{G}_{\mathbb{S}_{\boldsymbol{\phi}} = \mathbb{I}_{2\ell\times 2\ell}}$, identified through the approach described in section \ref{sec:Flow reconstruction using system identification and optimal estimation theory},
\begin{subequations}
\begin{equation}
\boldsymbol{x}[k+1] = \mathsfbi{A} \boldsymbol{x}[k] +\mathsfbi{K}\boldsymbol{e}[k],
\label{eq: plant2a}
\end{equation}
\mbox{with output}
\begin{equation}
 \boldsymbol{s}[k] =
\boldsymbol{\phi} \boldsymbol{a}[k] = \boldsymbol{\phi} \mathsfbi{C}  \boldsymbol{x}[k].
\label{eq: plant2b}
\end{equation}
\label{eq:_fully_sensed system}
\end{subequations}
Since the system is fully sensed, $ \boldsymbol{s}[k]=\left[ u'_1[k], u'_2[k], \dots, u'_\ell[k], v'_1[k],v'_2[k],\dots, v'_\ell[k]\right]^\top$.  If the $i$-th element $(i=1\dots m)$ of the input vector $\boldsymbol{e}[k]$ is a unit impulse, i.e.\ $\boldsymbol{e}_i[k]=\delta[k]$, it is easy to  show that the $j$-th element $(j=1\dots 2\ell)$ of the response (output) will be ${s}_{j}^{(i)}[k]=\left( \boldsymbol{\phi} \mathsfbi{C} \right)_j \mathsfbi{A}^{k-1} \mathsfbi{K}_i$, where $\left( \boldsymbol{\phi} \mathsfbi{C} \right)_j$ is the $j$-th row of $\boldsymbol{\phi} \mathsfbi{C}$ and $\mathsfbi{K}_i$ is the $i$-th column of $\mathsfbi{K}$, see \cite{Antoulas_2005}. The corresponding contribution to the fluctuating kinetic energy at instant $k$ will be $\frac{1}{2} \left({s}_{j}^{(i)}[k]\right)^2= \frac{1}{2}{s}_{j}^{(i)}[k]{s}_{j}^{(i)}[k]$ and the integral over time $\frac{1}{2}\sum_{k=0}^\infty \left({s}_{j}^{(i)}[k]{s}_{j}^{(i)}[k]\right)\Delta t$. Let's assume now that the unit impulse is applied successively to all $m$ elements of $\boldsymbol{e}[k]$ one by one. The integral of the total fluctuating kinetic energy over space and time can be written as:
\begin{equation}
\begin{split}
J &= \int_{0}^\infty \int_{Y} \int_{X} \frac{u'^2+v'^2}{2} \:\mathrm{d}x \:\mathrm{d}y \:\mathrm{d}t \approx \sum_{k=0}^\infty \left [\sum_{i=1}^m  \left( \sum_{j=1}^{2\ell}  \frac{{s}_j^{(i)}[k] {s}_j^{(i)}[k]}{2} \Delta x \Delta y \right) \right] \Delta t = \\ 
&\frac{\Delta x \Delta y \Delta t}{2} \sum_{k=0}^\infty \left [\sum_{i=1}^m  \left(   \boldsymbol{s}^{(i)}[k] \right)^\top \boldsymbol{s}^{(i)}[k] \right] 
= \frac{\Delta x \Delta y \Delta t}{2} 
\lVert \boldsymbol{G}_{\mathbb{S}_{\boldsymbol{\phi}}=\mathbb{I}_{2\ell\times 2\ell}} \rVert_2^2,
\end{split}
\label{eq: integral_TKE}
\end{equation}
where  $\boldsymbol{s}^{(i)}[k]=\left[ s_1^{(i)}[k], s_2^{(i)}[k], \dots, s_{2\ell}^{(i)}[k] \right ]^\top$ and  $\lVert \boldsymbol{G}_{\mathbb{S}_{\boldsymbol{\phi}}=\mathbb{I}_{2\ell\times 2\ell}} \rVert_2^2$ is the 
$\mathcal{H} _2$ norm of the fully sensed system:
\begin{equation}
\lVert \boldsymbol{G}_{\mathbb{S}_{\boldsymbol{\phi}}=\mathbb{I}_{2\ell\times 2\ell}} \rVert_2^2=\sum_{k=0}^\infty \mathrm{tr} \left ( \left(\boldsymbol{\phi} \mathsfbi{C} \right) \mathsfbi{A}^k \mathsfbi{K} \mathsfbi{K}^\top \left(\mathsfbi{A}^\top\right)^k  \left(\boldsymbol{\phi} \mathsfbi{C}\right)^\top\right) = \mathrm{tr} \left( \left(\boldsymbol{\phi} \mathsfbi{C} \right)  \boldsymbol{W}_c  \left ( \boldsymbol{\phi} \mathsfbi{C}\right)^\top \right),
    \label{eq: H2norm_fully_sensed}
\end{equation}
and $ \boldsymbol{W}_c$ is the controllability Gramian,
\begin{equation}
\boldsymbol{W}_c=\sum_{k=0}^\infty {\mathsfbi{A}}^k \mathsfbi{K} \mathsfbi{K}^\top \left({\mathsfbi{A}^\top}\right)^k.
    \label{eq: Contr_Gramian}
\end{equation}
The partially sensed system, $\boldsymbol{G}_{\mathbb{S}_{\boldsymbol{\phi}}}$, \begin{subequations}
\begin{equation}
    \boldsymbol{x}[k+1] = \mathsfbi{A} \boldsymbol{x}[k] + \mathsfbi{K}\boldsymbol{e}[k],
\end{equation}
\begin{equation}
\boldsymbol{s}[k] =
\mathbb{S}_{\boldsymbol{\phi}}\boldsymbol{\phi} \boldsymbol{a}[k] = \mathbb{S}_{\boldsymbol{\phi}}\boldsymbol{\phi} \mathsfbi{C}  \boldsymbol{x}[k],
\end{equation}
\label{eq:_partially_sensed system}
\end{subequations}
is characterised by the $p \times (2\ell) $ row-selection matrix ${\mathbb{S}_{\boldsymbol{\phi}}}$ and 
has $\mathcal{H}_2$ norm:
 \begin{equation}
\lVert \boldsymbol{G}_{\mathbb{S}_{\boldsymbol{\phi}}} \rVert_2^2 = \mathrm{tr} \left(\left(\mathbb{S}_{\boldsymbol{\phi}} \boldsymbol{\phi} \mathsfbi{C} \right)  \boldsymbol{W}_c  \left ( \mathbb{S}_{\boldsymbol{\phi}} \boldsymbol{\phi} \mathsfbi{C}\right)^\top \right)=\mathrm{tr} \left(\left(\mathbb{S}_{\boldsymbol{\phi}} \boldsymbol{\phi}  \right)  \boldsymbol{W}_{oc} \left ( \mathbb{S}_{\boldsymbol{\phi}} \boldsymbol{\phi} \right)^\top \right),
\label{eq: H2norm_partially_sensed}
\end{equation}
where $ \boldsymbol{W}_{oc}=\mathsfbi{C}  \boldsymbol{W}_c \mathsfbi{C}^\top$ is the output controlability Gramian,  which is independent of the coordinate transformation $T$ of the state vector.  

The objective now is to select  ${\mathbb{S}_{\boldsymbol{\phi}}}$ in a way that optimally preserves a measure of the $\mathcal{H}_2$ norm of the fully-sensed system. In other words, we aim to rank the sensors according to their contribution to the $\mathcal{H}_2$ norm, and keep only the ones that make the most significant contribution.  Different options are available, but the one that leads to a computationally tractable algorithm is to preserve the logarithm of the determinant of the matrix   $\left( \mathbb{S}_{\boldsymbol{\phi}} \boldsymbol{\phi}  \right)  \boldsymbol{W}_{oc}  \left ( \mathbb{S}_{\boldsymbol{\phi}} \boldsymbol{\phi} \right)^\top$, thus
\begin{equation}
\mathbb{S}_{\boldsymbol{\phi}\star} \approx \underset{\mathbb{S}_{\boldsymbol{\phi}}}{\textup{argmax}}  \: \log \left\{ \det \left[  \mathbb{S}_{\boldsymbol{\phi}} \boldsymbol{\phi} \boldsymbol{W}_{oc}  \boldsymbol{\phi}^\top \mathbb{S}_{\boldsymbol{\phi}}^\top \right] \right \}
\label{eq: submatrix volume maximisation_1}
\end{equation}
For $n \ge m$, matrix $\boldsymbol{W}_{oc}$ is symmetric positive definite (otherwise it is rank deficient, thus semi-definite), and we can apply Cholesky decomposition, $ \boldsymbol{W}_{oc}=\mathsfbi{F} \mathsfbi{F}^\top$. Substituting in \eqref{eq: submatrix volume maximisation_1} we get
\begin{equation}
\mathbb{S}_{\boldsymbol{\phi}\star} \approx \underset{\mathbb{S}_{\boldsymbol{\phi}}}{\textup{argmax}}  \: \log \left\{ \det \left[  \mathbb{S}_{\boldsymbol{\phi}} \boldsymbol{\phi} \mathsfbi{F} \mathsfbi{F}^\top  \boldsymbol{\phi}^\top \mathbb{S}_{\boldsymbol{\phi}}^\top \right] \right \}=
\underset{\mathbb{S}_{\boldsymbol{\phi}}}{\textup{argmax}}  \: \log \left\{ \det \left[  \mathbb{S}_{\boldsymbol{\phi}} \boldsymbol{\phi} \mathsfbi{F} 
\left(\mathbb{S}_{\boldsymbol{\phi}}    \boldsymbol{\phi}  \mathsfbi{F} \right)^\top \right] \right \}.
\label{eq: submatrix volume maximisation_2}
\end{equation}
Taking the number of sensors equal to the number of retained modes i.e.\ $p=m$, matrix  $\mathbb{S}_{\boldsymbol{\phi}} \boldsymbol{\phi} \mathsfbi{F}$ becomes square, and using the multiplicative property of determinants, the term within the curly brackets becomes $\det \left[ \left( \mathbb{S}_{\boldsymbol{\phi}} \boldsymbol{\phi} \mathsfbi{F} \right) \left ( \mathbb{S}_{\boldsymbol{\phi}} \boldsymbol{\phi} \mathsfbi{F}\right)^\top \right]=\left( \det \mathbb{S}_{\boldsymbol{\phi}} \boldsymbol{\phi} \mathsfbi{F} \right)^2$, thus
\begin{equation}
\mathbb{S}_{\boldsymbol{\phi}\star} \approx \underset{\mathbb{S}_{\boldsymbol{\phi}}}{\textup{argmax}} \left\{ 2 \mathrm{log} \left | \det \left( \mathbb{S}_{\boldsymbol{\phi}} \boldsymbol{\phi}  \mathsfbi{F} \right) \right | \right \}.
\label{eq: submatrix volume maximisation_3}
\end{equation}
Since the logarithm is a monotonic function of its argument, this is equivalent to:
\begin{equation}
\mathbb{S}_{\boldsymbol{\phi}\star} \approx \underset{\mathbb{S}_{\boldsymbol{\phi}}}{\textup{argmax}} \left | \det \left( \mathbb{S}_{\boldsymbol{\phi}} \boldsymbol{\phi}  \mathsfbi{F} \right) \right |,
\label{eq: submatrix volume maximisation}
\end{equation}
\noindent where $\mathbb{S}_{\boldsymbol{\phi}\star}$ is the optimal row-selection matrix that maximises the absolute value of the determinant of $\mathbb{S}_{\boldsymbol{\phi}} \boldsymbol{\phi}  \mathsfbi{F}$. Matrix $\mathbb{S}_{\boldsymbol{\phi}\star}$ contains either 1 or 0 entries; the former mark the locations of the selected sensors. 

Equation \eqref{eq: submatrix volume maximisation} defines a sub-matrix volume optimisation problem that can be solved by QR factorisation of the matrix $(\boldsymbol{\phi}\mathsfbi{F})^\top$. This operation will provide the permutation matrix $\mathsfbi{P}$ that satisfies 
\begin{equation}
\left (\boldsymbol{\phi}\mathsfbi{F} \right )^{\top} \mathsfbi{P} = \mathsfbi{Q} \mathsfbi{R},
    \label{eq: QR pivoting}
\end{equation}
\noindent where $\mathsfbi{Q}$ is a unitary matrix and $\mathsfbi{R}$ is an upper-triangular, diagonally-dominant matrix, with the diagonal elements ranked in descending order i.e.\ $\left | R_{11} \right | \geq \left | R_{22} \right | \geq \left | R_{33} \right |$ etc. (these matrices should not be confused with the covariance matrices $\mathsfbi{Q}$ and $\mathsfbi{R}$ defined in section \ref{sec:Flow reconstruction using system identification and optimal estimation theory}; we expect that the distinction will be clear from the context). The permutation matrix $\mathsfbi{P}$ in \eqref{eq: QR pivoting} stores the pivot indices of the selected columns of $\left (\boldsymbol{\phi}\mathsfbi{F} \right )^{\top}$. The absolute value of the determinant of the row-selected submatrix can be written as, 
\begin{equation}
\left | \det \left (\boldsymbol{\phi}\mathsfbi{F} \right )^{\top} \mathsfbi{P}_{:,1:p} \right | = \left | \det \mathsfbi{Q} \right | \left | \det \mathsfbi{R}_{:,1:p}  \right | = \prod_{i=1}^{p} \left | R_{ii} \right |,
    \label{eq: absolute det}
\end{equation}
which suggests that the descending diagonal elements of  $\mathsfbi{R}$ result in optimal submatrix determinants. Therefore, the leading $p$ columns of $\mathsfbi{P}$ define the optimal row-selection matrix,
\begin{equation}
\mathbb{S}_{\boldsymbol{\phi}\star} = \mathsfbi{P}_{:,1:p}.
    \label{eq: leading columns of P}
\end{equation}
It should be emphasised that each row of $\mathbb{S}_{\boldsymbol{\phi}\star}$ defines both the physical location of the sensor as well as the velocity component, $u'$ or $v'$, that needs to be measured.  

The identified sensors are sub-optimal for two reasons. First, they are obtained by considering the identified open-loop system \eqref{eq: plant} and not the estimator system \eqref{eq: kalman filter formulation}. Strictly speaking, the sensors should minimise the $\mathcal{H}_2$ norm of the system that maps the process and output noise, $\boldsymbol{w}[k]$ and $\boldsymbol{v}[k]$, to the estimation error $\mathbb{S}_{\boldsymbol{\phi}}\mathsfbi{\phi}\mathsfbi{C} \left( \hat{\boldsymbol{x}}[k]- \boldsymbol{x}[k]\right)$ (details about this system and its $\mathcal{H}_2$ norm can be found in chapter 5 of \cite{GreenLimebeer95}). This is for example the approach taken by \cite{Chen_Rowley_2011} for optimal sensor and actuation placement to  control the 1D Ginzburg-Landau equation. This approach however results in an  optimisation problem which is difficult to solve (for example one needs to start from different initial sensor locations to avoid local minima). Second, the QR decomposition maximises the determinant of  $\left( \mathbb{S}_{\boldsymbol{\phi}} \boldsymbol{\phi} \mathsfbi{F} \right) \left ( \mathbb{S}_{\boldsymbol{\phi}} \boldsymbol{\phi} \mathsfbi{F}\right)^\top$, not the trace, see  \eqref{eq: H2norm_partially_sensed}. Nevertheless the approach employed in this paper is computationally efficient and, as will be seen later, works quite well. Performance can be improved if the identified locations are used as initial conditions to accelerate the convergence of the iterative approach of \cite{Chen_Rowley_2011}. 

In the following section \S\ref{sec:Performance assessment of dynamic estimation} we compare the results from three flow reconstruction methods, namely stochastic system identification with Kalman filtering (section \ref{sec:Flow reconstruction using system identification and optimal estimation theory}, fig. \ref{fig:kalman filter}), direct identification (section \ref{sec:Direct identification of the dynamical estimator}, fig. \ref{fig:original estimator}) and static reconstruction. For the former method, we extract velocity measurements from sensors placed at the peaks of the POD eigenmodes (see  \citet{Yildirim}) as well as at the QR locations. At the POD peaks both velocity fluctuations are measured, because the peak position is determined by calculating $\sqrt{\phi^{(u')}(\boldsymbol{x})^2 + \phi^{(v')}(\boldsymbol{x})^2}$. Hence, the number of measurements is twice the number of sensor locations. On the other hand, either $u'$ or $v'$ is measured at the QR sensors, so the number of signals is equal to the number of sensors. Below we compare the performance of the two approaches for the same number of measurements. Finally, for the latter two methods, direct identification and static reconstruction, we consider sensor locations obtained from the "static" version of the QR pivoting algorithm, i.e.\  $\boldsymbol{\phi}^{\top} \mathsfbi{P} = \mathsfbi{Q} \mathsfbi{R}$, see \citet{Manohar2}.

\section{Assessment of reconstruction performance\label{sec:Performance assessment of dynamic estimation}}

\subsection{Sensor locations} \label{subsec:Sensor_locations}
\begin{figure}[ht]
    \centering
    \centerline{\includegraphics[width=\textwidth]{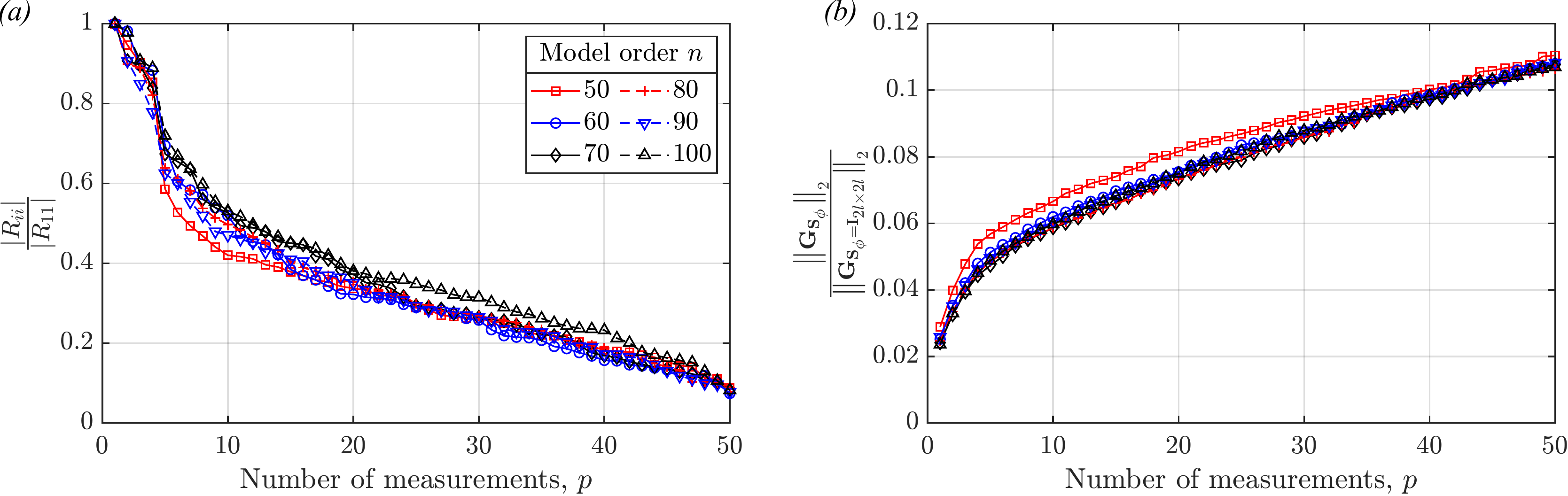}}
    \caption{($a$) Normalised distribution of the diagonal elements of the $\mathbf{R}$ matrix and ($b$) $\mathcal{H}_2$ norm of the partially observed system, $\lVert \boldsymbol{G}_{\mathbb{S}_{\boldsymbol{\phi}}} \rVert_2$, expressed as a fraction of the $\mathcal{H}_2$ norm of the fully observed system, $\lVert \boldsymbol{G}_{\mathbb{S}_{\boldsymbol{\phi}=\mathbb{I}_{2\ell \times 2\ell}}} \rVert_2$. Results are shown for different model orders.}
    \label{fig:Rii_H2}
\end{figure}

We start with examining the sensor locations found by the QR algorithm. In fig. \ref{fig:Rii_H2}$a$, the magnitude of the (normalised) diagonal elements of matrix $\mathsfbi{R}$, equation \eqref{eq: QR pivoting}, are plotted against the number of measurements $p$ (equal to the number of sensors) for different model orders, $n$ (the latter dictates the maximum value of $p$). The magnitude of $\left | R_{ii} \right |$ is an indicator of the importance of the sensor; higher values indicate more valuable sensors. The shape of the curves is similar for all $n$; rapid drop of $\left | R_{ii} \right |$ is followed by slower decay for $p$ greater than 7-8. This is corroborated by fig. \ref{fig:Rii_H2}$b$ which plots the $\mathcal{H}_2$ norm of the partially observed system $\boldsymbol{G}_{\mathbb{S}_{\boldsymbol{\phi}}}$ (normalised with the norm of the fully sensed system) against $p$. The first 7-8 sensors contribute to the steepest growth of $\lVert \boldsymbol{G}_{\mathbb{S}_{\boldsymbol{\Phi}}} \rVert_2$,  while for larger $p$ the slope is reduced. Interestingly, for $n>50$ the curves in fig. \ref{fig:Rii_H2}$b$ collapse well with small scatter. The scatter is higher in fig. \ref{fig:Rii_H2}$a$. 

\begin{figure}[ht]
  \centerline{\includegraphics[width=0.65\textwidth]{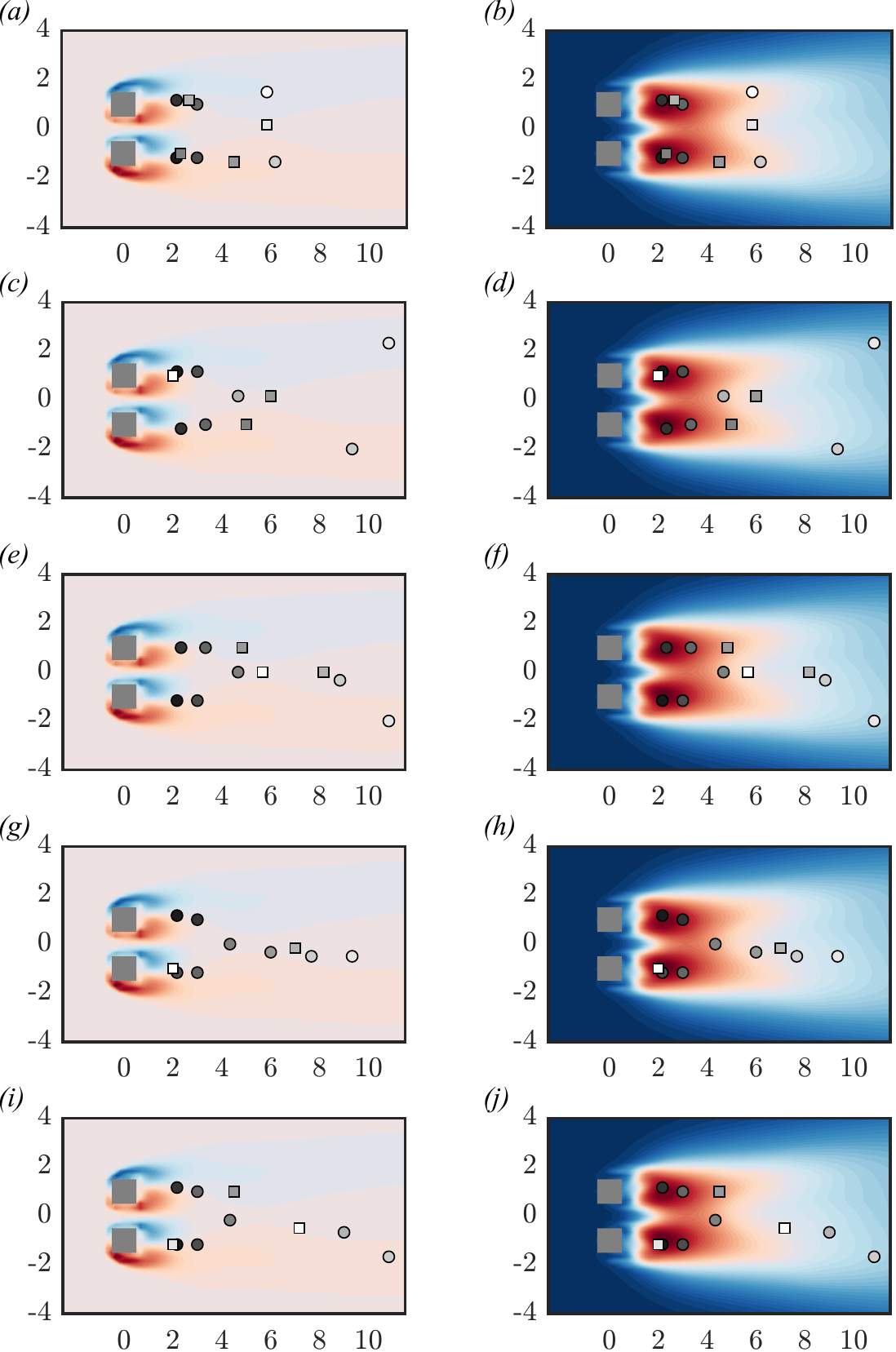}}
  \caption{Locations of the leading 10 sensors identified by the application of the QR pivoting algorithm on $(\boldsymbol{\phi}\mathsfbi{F})^\top$ for model orders $n=50$ ($a$-$b$), $n=70$ ($c$-$d$), $n=80$ ($e$-$f$), $n=90$ ($g$-$h$) and $n=100$ ($i$-$j$). Sensors are superimposed on time-averaged vorticity (left column) and fluctuating kinetic energy contours (right column). A grey scale is used to indicate the sensors ranking from most (darker) to least (lighter) important. Square $(\square)$ and circular $(\bigcirc)$ markers denote measurements of $u'$ and $v'$ velocity components, respectively.}
\label{fig:convergenceQR}
\end{figure}

The locations of the leading 10 sensors for different model orders $n$ are shown in fig. \ref{fig:convergenceQR}. The results are superimposed on contours of time-average vorticity (left column) and fluctuating kinetic energy (right column).  The employed grey scale, from dark to light, ranks the sensors from most to least important respectively. Square and circular markers indicate $u'$ and $v'$ signals respectively. For $n=50$ (first row at the top), the 2 most important sensors are placed behind each cylinder (at $x \approx 2$) at the region of highest fluctuating kinetic energy (see right column) which makes intuitive sense. Both record the cross-stream velocity component. Increasing the model order, $n$, preserves the location of the leading 2 sensors and readjusts the placement of the others. The coordinates of the dominant 6 sensors have converged at $n=90$ (panels g and h); all are located in the near wake, $x<5$, 5 are behind the cylinders and 1 in the centreline. It is also interesting to notice that 5 out of 6 sensors measure the $v'$ velocity component, and only 1 measures the $u'$ component (and it is collocated with a sensor that also measures $v'$ behind the bottom cylinder). For $n=100$ (bottom row) the most downstream sensors have $y<0$; this does not imply that the method results in asymmetrically placed sensors, symmetry is restored for larger $p$ as will be seen later. 

\begin{figure}
    \centering
    \includegraphics[width=0.9\linewidth]{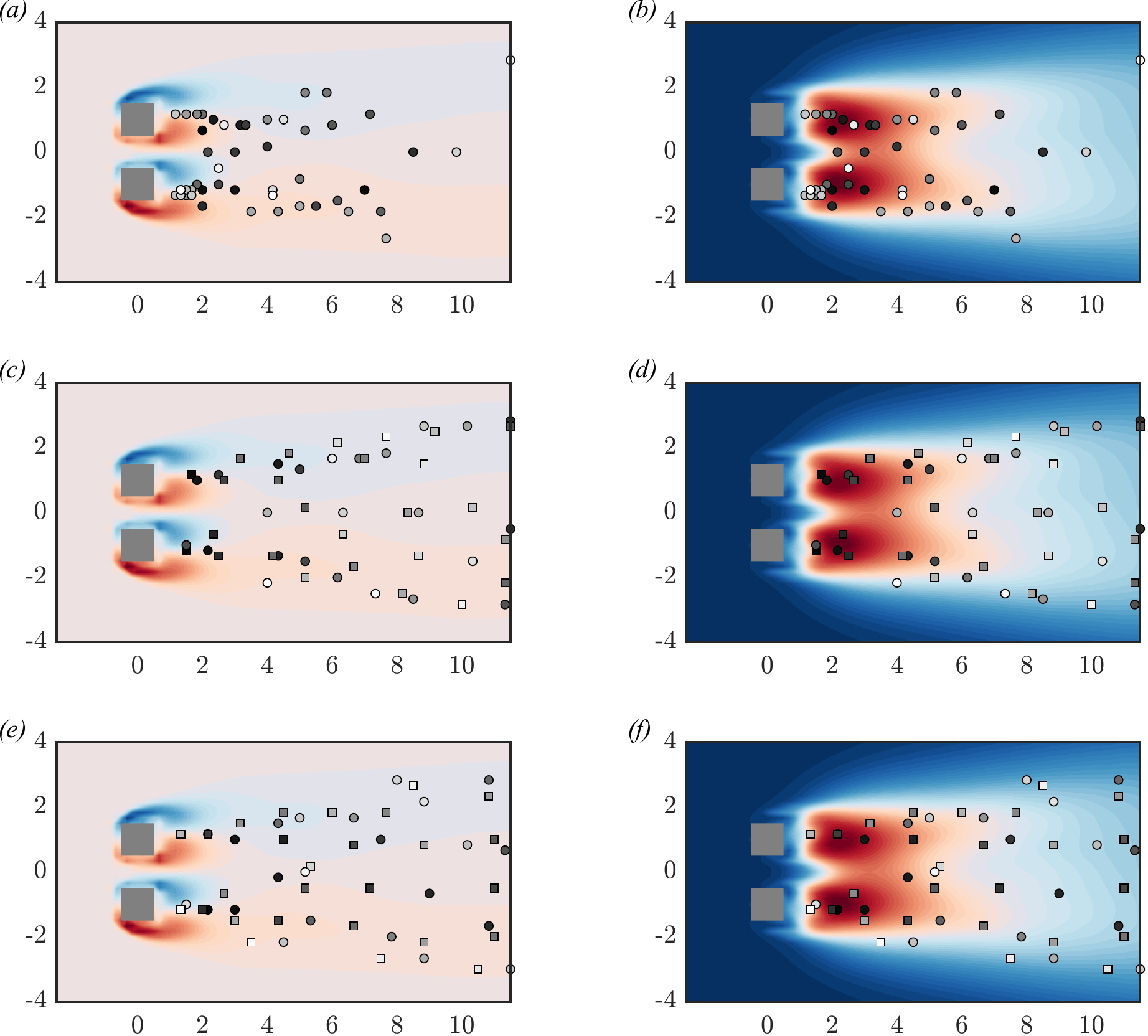}
    \caption{Sensors placed at the POD peaks (top row, $a$-$b$), computed from the QR pivoting algorithm on the POD matrix $\boldsymbol{\phi}^\top$ (middle row, $c$-$d$) and on the matrix $(\boldsymbol{\phi} \mathsfbi{F})^\top$ for model order $n=100$ (bottom row, $e$-$f$). Sensor locations are superimposed on contours of time-averaged vorticity (left column) and fluctuating kinetic energy (right column). A grey scale is used to indicate the sensor ranking, from most (darker) to least (lighter) important. Square $(\square)$ and circular $(\bigcirc)$ markers denote $u'$ and $v'$ measurements respectively in the QR sensor libraries. The POD peaks are shown as circular markers, but both velocity fluctuations are measured.}
    \label{fig:sensors_comparison}
\end{figure}

In fig. \ref{fig:sensors_comparison} we compare the spatial distribution of sensors located at the POD peaks (top row) against the distribution computed from "static" (middle row) and "dynamic" (bottom row) QR pivoting. The time-average vorticity contours (left column) indicate the presence of two spreading wakes, one behind each cylinder. On average, the QR sensors are evenly spread within each wake (see panels (c) and (e)), whereas the POD sensors are more densely clustered behind the cylinders, see panel (a). This is corroborated by inspecting panel (b) which shows that the sensors mainly occupy the region of high TKE, with only very few sitting outside this region. On the other hand, a significant number of low ranking QR sensors are within this area, as can be seen in panels (d) and (f).   

\subsection{Stability of the identified models \label{subsubsec:Stability_of_identified_models}}
\begin{figure}[ht]
    \centerline{\includegraphics[width=0.9\textwidth]{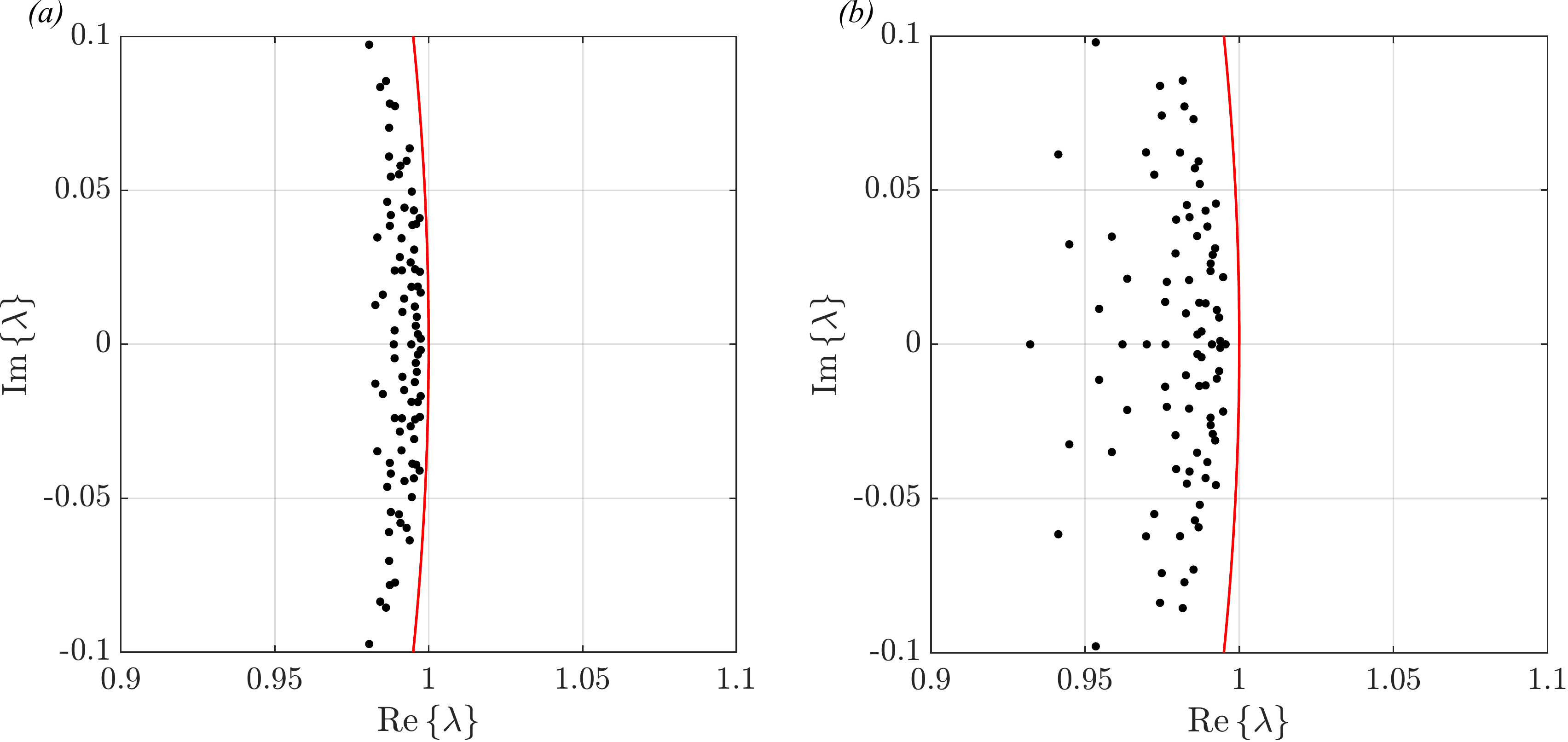}}
    \caption{Eigenvalue spectrum of ($a$) the identified matrix $\mathsfbi{A}$ and ($b$) the Kalman filter matrix $\mathcal{A}=\mathsfbi{A}-\mathcal{L}\mathsfbi{SC}$. The stability boundary (unit circle) is shown in red. The model order is $n=90$ and the Kalman filter is derived using the leading 10 QR sensors, i.e.\ $p=10$.}
    \label{fig:eigenvalues plant-model}
\end{figure}

Before investigating the statistics of the reconstructed velocity fields, we first analyse the stability of the identified model \eqref{eq: plant innovation form} and that of the Kalman filter estimator \eqref{eq: kalman filter formulation}. Panel (a) of figure \ref{fig:eigenvalues plant-model} shows the  eigenvalues of the identified matrix $\mathsfbi{A}$; all of them are within the unit circle indicating that the model is stable. The fact that they are clustered near the edge of the stable region indicates that they are very lightly damped, as expected from the self-sustained vortex shedding process. The eigenvalues of the matrix  $\mathcal{A}=\mathsfbi{A}-\mathcal{L}\mathsfbi{SC}$ of the dynamic estimator are shown in panel (b).  The spectrum is different compared to that of $\mathsfbi{A}$, and the eigenvalues are slightly further away from the unit circle (indicating enhanced stability). 

\begin{figure}[ht]
    \centerline{\includegraphics[width=0.9\textwidth]{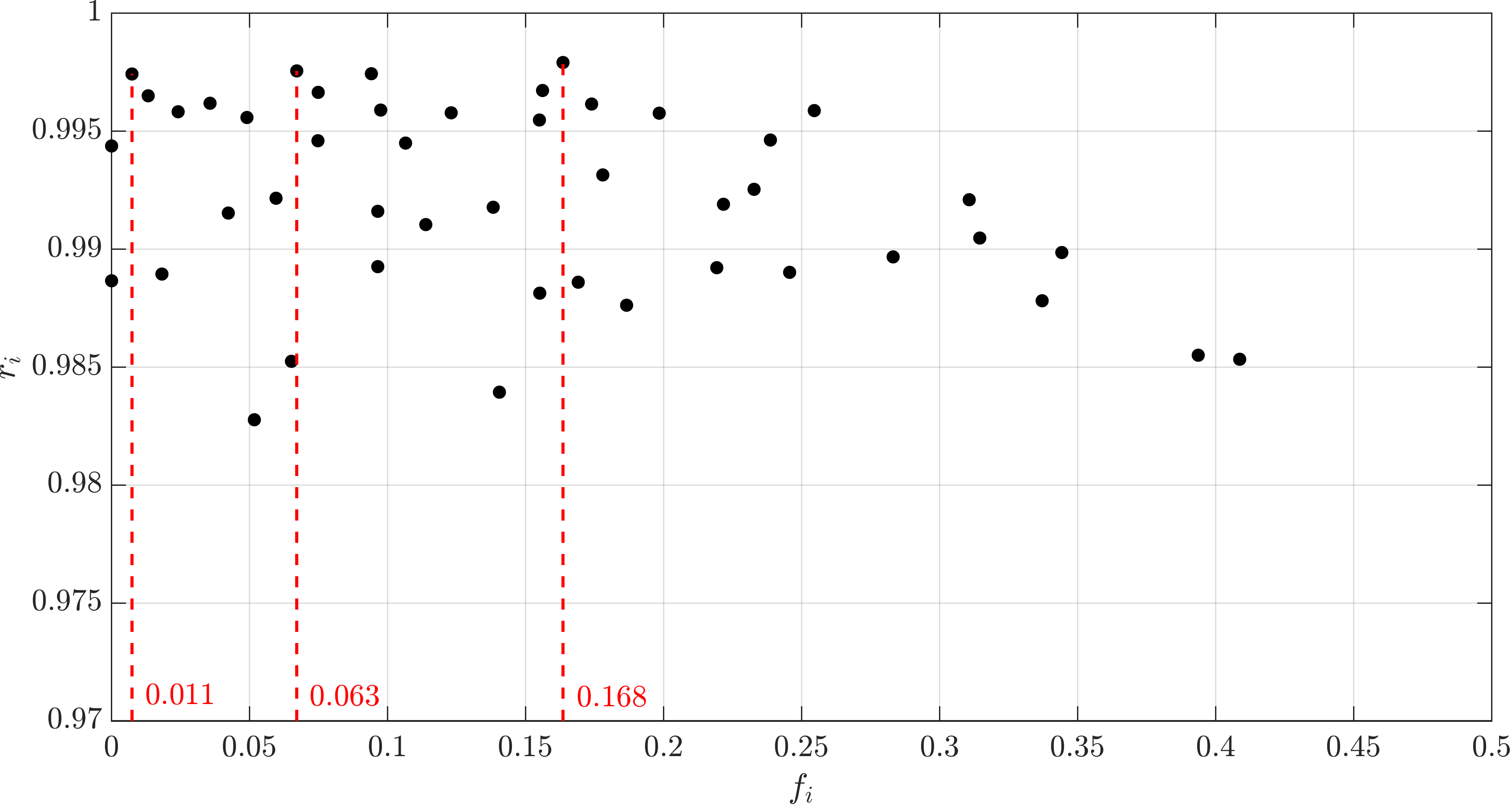}}
    \caption{Magnitude $r_i$ and frequency $f_i$ of the eigenvalues of matrix $\mathsfbi{A}$ for model order $n=90$. Only eigenvalues with positive frequencies are shown.}
    \label{fig:frequencies plant}
\end{figure}

The magnitudes $r_i$ and frequencies $f_i$ of the complex eigenvalues of matrix $\mathsfbi{A}$ are plotted in figure \ref{fig:frequencies plant}. The frequencies are obtained from $f_i=\phi_i/(2 \pi \Delta t)$, where $\phi_i$ is the phase of the complex eigenvalue and $\Delta t$ is the step size between successive time instants (here $\Delta t=0.04)$. 
The eigenvalues closest to the stability boundary $r = 1$ are marked with vertical red-dashed lines and are associated with the dominant large-scale motions of the flow. More specifically, the identified frequency 0.063 is present in the first POD mode pair (see figure \ref{fig:modes}) and is associated with the flapping motion of the gap flow; the frequency 0.168 corresponds to the main shedding event; and the third frequency 0.011 appears in the fourth POD mode pair. 

\subsection{Reconstruction of velocity statistics\label{subsubsec:Reconstruction of velocity statistics}}
The performance of the dynamic estimators sketched in figures \ref{fig:kalman filter} and \ref{fig:original estimator} is tested systematically by investigating the effect of the model order, $n$, and the number of sensors, $p$. More specifically, we employ two performance indicators, the scalar $\textup{FIT}\:[\%]$ defined as, 
\begin{equation}
	\textup{FIT}\: [\%] = 100 \left (1 - \frac{\int   \overline{\left(u_i(\boldsymbol{x},t)-\hat{u}_i(\boldsymbol{x},t)\right)^2} \textup{d}\boldsymbol{x} }{\int \overline{u_{i}^{2}(\boldsymbol{x},t)}\textup{d}\boldsymbol{x}} \right ),
	\label{eq:TKE_FIT}
\end{equation} 
and the field $\textup{FIT}(\boldsymbol{x})\:[\%]$ defined as,
\begin{equation}
		\textup{FIT}(\boldsymbol{x})\: [\%] = 100 \left (1 -  \frac{  \overline{(u_i(\boldsymbol{x},t)-\hat{u}_i(\boldsymbol{x},t))^2} }{\overline{u_i^2(\boldsymbol{x},t)}} \right ).
	\label{eq:TKE_FIT_field}
\end{equation}

\begin{figure}[ht]
    \includegraphics[width=\textwidth]{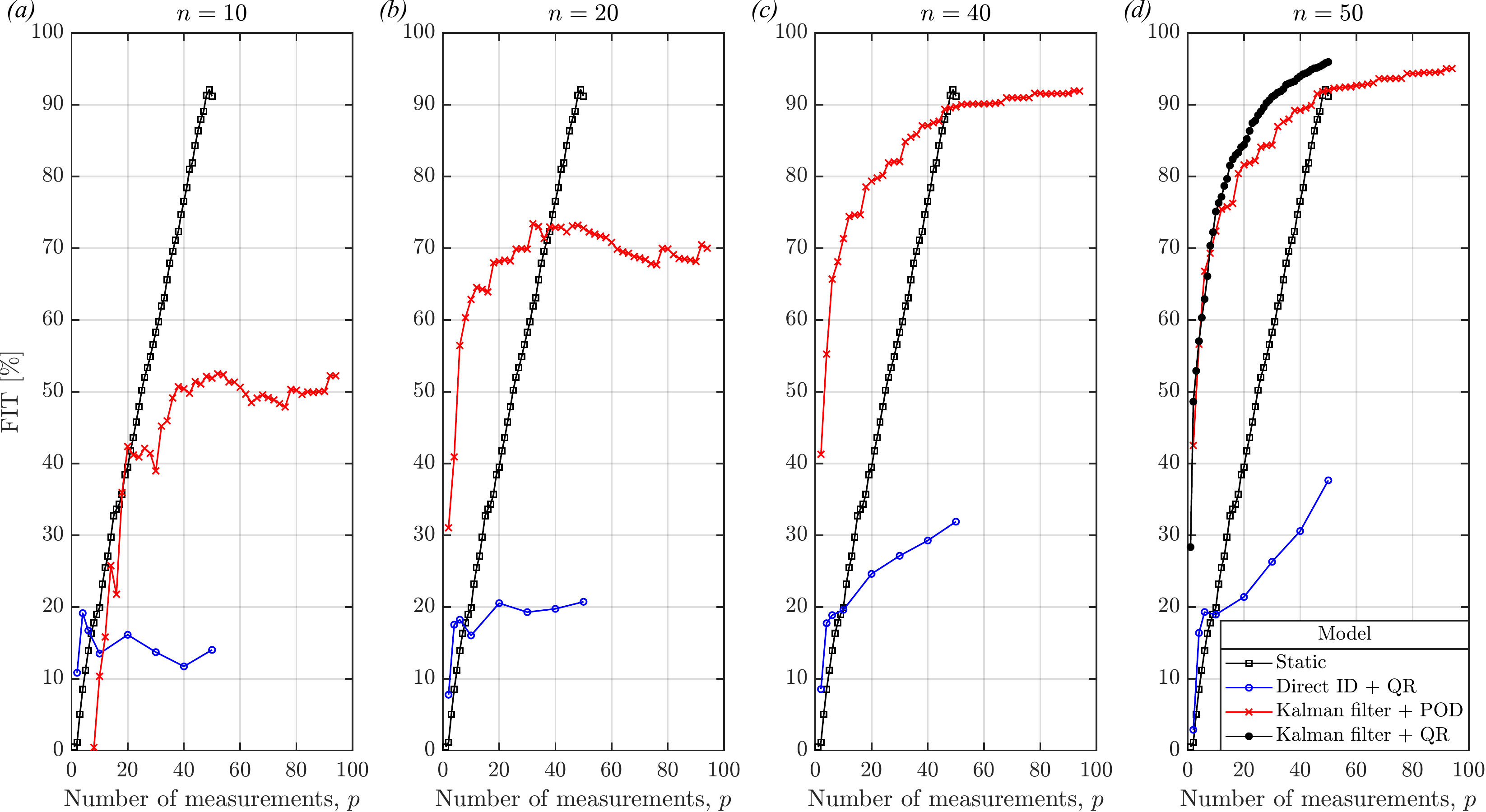} \par
    \vspace{0.5cm}
    \includegraphics[width=\textwidth]{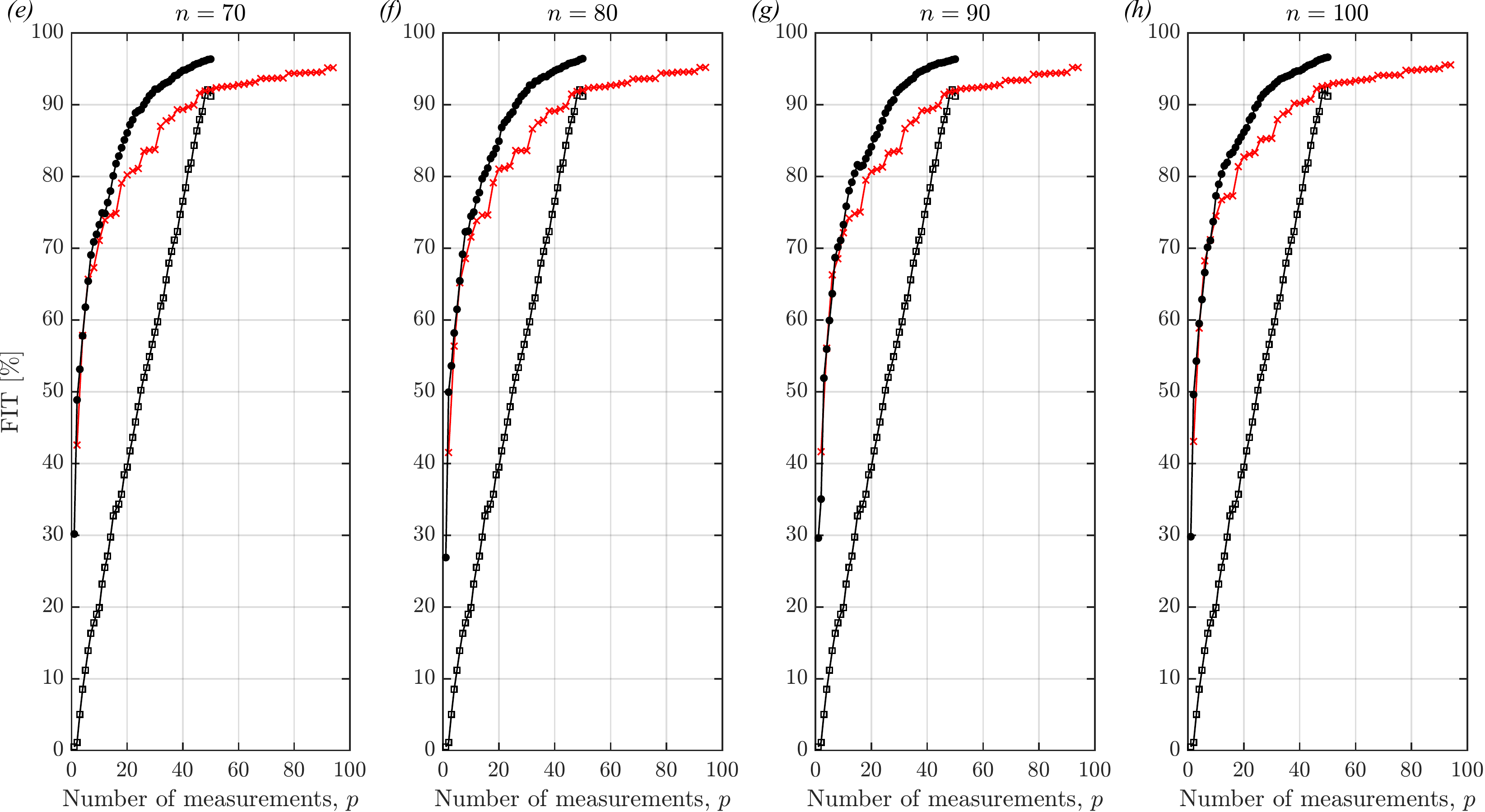}
    \caption{Scalar $\textup{FIT}\:[\%]$ against the number of measurements, $p$, for different model orders $n$, using static reconstruction (open black squares), direct identification with QR sensors (open blue circles), Kalman filter with POD sensors (red  crosses) and Kalman filter with QR sensors (solid black circles).}
    \label{fig:FIT curves}
\end{figure}

The scalar $\textup{FIT}\:[\%]$ against the number of measurements, $p$, for different model orders, $n$, is plotted in fig. \ref{fig:FIT curves}. The results from direct identification, marked by open blue circles in panels (a)-(d), are shown up to model order $n=50$ (we could not get results for larger $n$ using the \texttt{n4sid} command of \texttt{MATLAB}). The results display two main features. First, increasing the model order $n$ results in more accurate reconstruction; this is expected from the complexity of the underlying system dynamics. Second, performance improves monotonically with increasing $p$, but this necessitates $n=40,50$. The maximum  $\textup{FIT}\:[\%]$ reaches almost $40\%$; this relatively low value is most likely due to the fact that significant information, such as the system \& measurement noise and the shape of the POD modes, is not incorporated into the formulation of the estimator. 

The static reconstruction procedure implemented is identical to that of \citet{Manohar2}. The POD temporal coefficients are obtained from,
\begin{equation}
    \boldsymbol{\hat{a}}[k] = \mathsfbi{S}^{\dagger} \boldsymbol{s}[k],
    \label{eq:temporal_static}
\end{equation}
\noindent where the symbol $\dagger$ denotes the Moore-Penrose pseudo-inverse. Matrix $\mathsfbi{S}$ contains the rows of the matrix $\boldsymbol{\phi}$ that correspond to the location and type of measurement. Expression \eqref{eq:temporal_static} is obtained by ignoring the noise term in \eqref{eq: input with noise} and  pseudo-inverting.  The results in fig. \ref{fig:FIT curves}, marked with open black squares, are independent of the model order $n$ but are repeated in each plot to facilitate comparison. It is clear that the reconstruction quality improves steadily (and linearly) as $p$ increases. With $p=50$ QR sensors the scalar $\textup{FIT}\:[\%]$ reaches 91\%. When few measurements are considered, say $p<5-6$, static reconstruction under-performs with respect to direct identification. This indicates that dynamic estimation offers some benefits when the available sensor budget is small. On the other hand, provided we can afford a large number of measurements, static estimation becomes an accurate and computationally cheap method. 

Fig. \ref{fig:FIT curves} demonstrates clearly that identifying the underlying system dynamics first and then applying Kalman filter to obtain the dynamic estimator results in significantly improved reconstruction quality, see red crosses (for POD sensors) and filled circles (for QR sensors). Results with QR sensors are shown only for $n\ge m(=50)$. The FIT values become approximately independent of the model order for $n \ge 50$, hence a minimum model order, approximately equal to the number of retained POD modes, is required to obtain robust results. The two types of sensors have very similar performance for $1<p<10$. The highest increase of FIT is observed for low $p$ values, consistent with the behaviour shown earlier in figure \ref{fig:Rii_H2}. As $p$ increases, we get diminishing returns in the growth of FIT, a property called sub-modularity, see \citet{Summers_et_al_2016, Tzoumas_et_al_2016}. It is also clear that for $p>10$ the QR sensors consistently outperform the POD sensors for all model orders. 

\begin{figure} [ht]
    \centerline{\includegraphics[width=\textwidth]{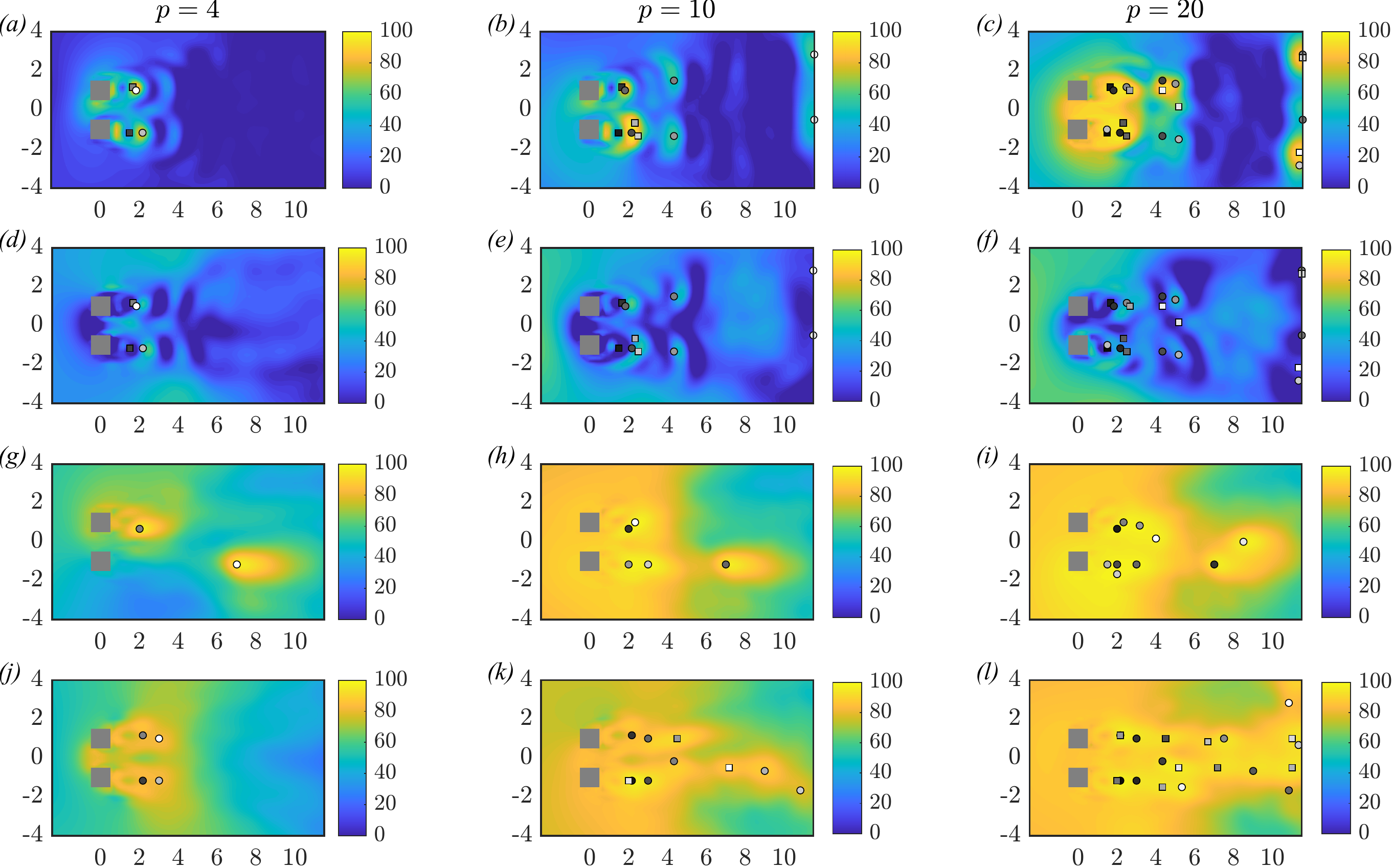}}
    \caption{Maps of $\textup{FIT}(\boldsymbol{x})$. ($a$-$c$) Static reconstruction using QR sensors based on the POD basis $\boldsymbol{\phi}$, ($d$-$f$) direct identification of order $n=50$ using QR sensors based on the POD basis $\boldsymbol{\phi}$, ($g$-$i$) Kalman filter of order $n=100$ using sensors at the POD peaks and ($j$-$l$) Kalman filter of order $n=100$ using QR sensors based on matrix $\boldsymbol{\phi}\mathsfbi{F}$. From left to right columns, the number of sensor measurements are $p=4$, $p=10$ and $p=20$. Square $(\square)$ and circular $(\bigcirc)$ markers indicate $u'$ and $v'$ measurements, respectively. The POD peaks are shown as circular markers, but both velocity fluctuations are measured.}
    \label{fig:FIT maps}
\end{figure}

From the scalar $\textup{FIT}\:[\%]$ alone we cannot get a clear understanding of the local reconstruction quality. Contours of $\textup{FIT}(\boldsymbol{x})$, defined in \eqref{eq:TKE_FIT_field}, are plotted in figure \ref{fig:FIT maps}. For the direct identification we chose $n=50$ (the maximum model order we could get results), while for the Kalman filter $n=100$ (for both POD and QR sensors). For $p=4$ (left column), static reconstruction and direct identification, panels (a) and (d) respectively, have limited accuracy practically everywhere behind the cylinders. The Kalman filters, panels (g) and (j), exhibit very high accuracy  around the sensor locations (higher than 80\%), and a large portion of the domain is reconstructed with 40-60\% accuracy.  With $p=10$ (middle column) the results from static reconstruction and direct identification, panels (b) and (e) respectively, show modest improvements. On the other hand, the Kalman filters, panels (h) and (k), exhibit a sharp increase in accuracy everywhere in the domain. Again the flow is almost accurately reproduced around the sensors. Regions of lower accuracy are away the measurement area, for example the top right of the domain in panel (h) and the patches above and below the high-accuracy horizontal strip in panel (k). With $p=20$ (right column), static reconstruction, panel (c), shows marked improvement in the near wake, but for direct identification, panel (f), this is not the case,  although the domain is more densely sensed. The Kalman filters, panels (i) and (l), significantly outperform the other two models. The contours indicate average reconstruction accuracy of 80-100\%. Furthermore, QR sensors (panel (l)) are seen to yield slightly better results compared to POD peaks (panel (i)). This is probably due to the broader distribution of the former, that allows for a more uniform sensing of the domain.

\begin{figure}
    \centerline{\includegraphics[width=\textwidth]{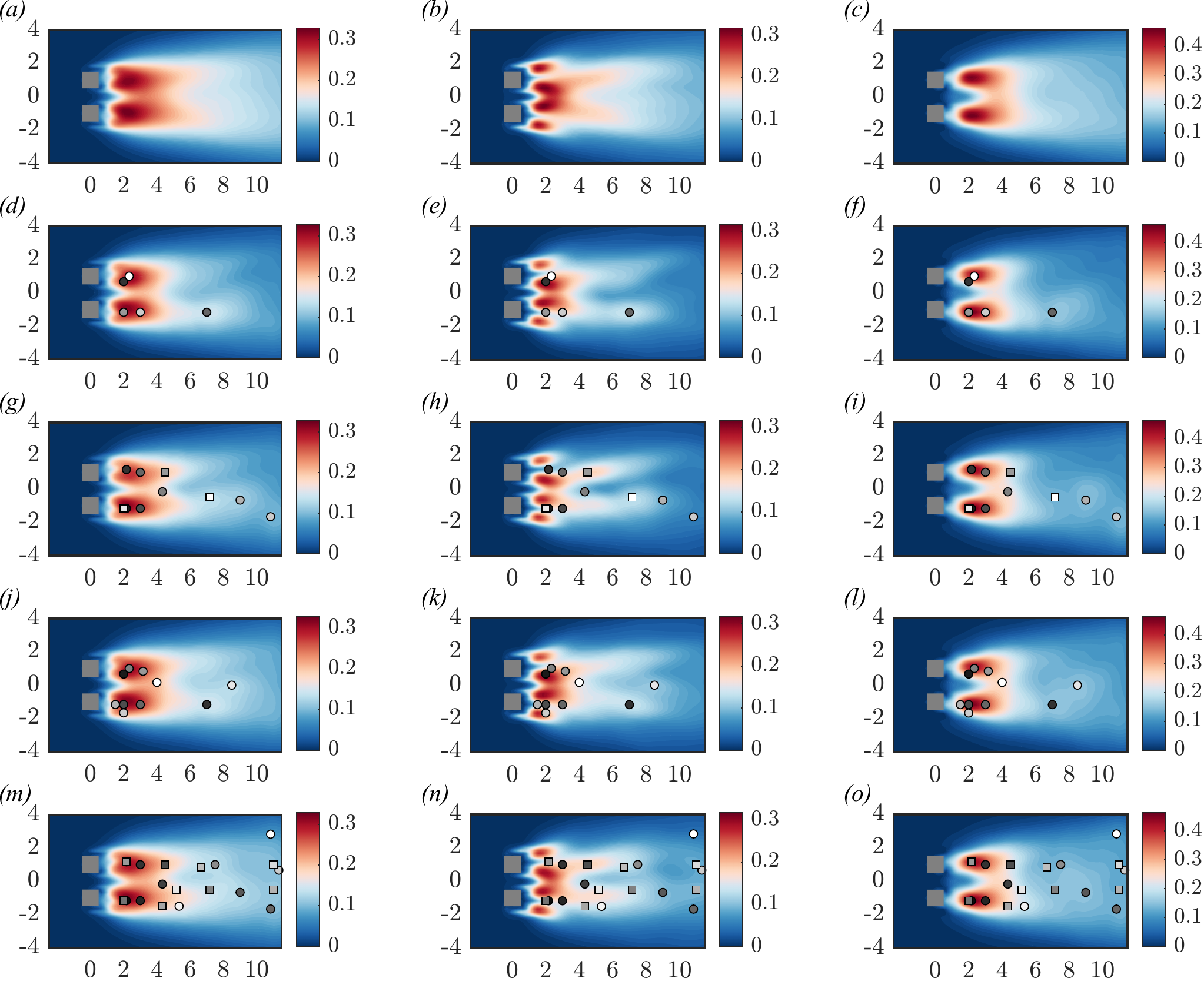}}
    \caption{Contour plots of velocity statistics: (left column)  fluctuating kinetic energy $k$, (middle column) $\overline{u'^2}$, and (right column) $\overline{v'^2}$. DNS data ($a$-$c$, top row), reconstruction using Kalman filter of order $n=100$ with ($d$-$f$, second row) $p=10$ POD measurements, ($g$-$i$, third row) $p=10$ QR measurements, ($j$-$l$, fourth row) $p=20$ POD measurements and ($m$-$o$, fifth row) $p=20$ QR measurements. Square $(\square)$ and circular $(\bigcirc)$ markers indicate $u'$ and $v'$ measurements respectively. POD peaks are indicated with circular markers, but both velocity components are recorded.} 
    \label{fig:statistics reconstruction}
\end{figure}

In figure \ref{fig:statistics reconstruction} we compare the reconstructed fields of the fluctuating kinetic energy $k$ and  Reynolds stresses $\overline{u'^2}$, $\overline{v'^2}$ (left, middle and right columns respectively). More specifically, we compare two sensor placement strategies, QR and POD (both with Kalman filter of order $n=100$) with $p=10$ and $p=20$ sensors. The top row, panels (a)-(c), shows the true fields obtained directly from the full DNS data. The near wake regions behind the cylinders display the highest fluctuations and are symmetric with respect to the centreline line $y=0$, as expected. The Kalman filter is able to reproduce these regions very well with only $p=10$ measurements, as can be seen from the second and third rows, panels (d)-(i). There are two regions of large $\overline{u'^2}$ (above the top cylinder and below the bottom) where no sensor point is placed, but still the reconstruction is very good. With $p=10$, the QR sensors, panel (h), resolve slightly better the two stretched legs of large $\overline{u'^2}$ compared to the POD sensors, panel (e), while minor differences are visible in the $\overline{v'^2}$ contours. When the number of measurements is increased to $p=20$, there is  modest improvement with the POD sensors, fourth row panels (j)-(l), and a more significant improvement with the QR sensors, fifth row panels (m)-(o). This observation agrees with the results presented previously, in that QR sensors perform best for $p\geq20$.

\begin{figure}[ht]
    \centerline{\includegraphics[width=\textwidth]{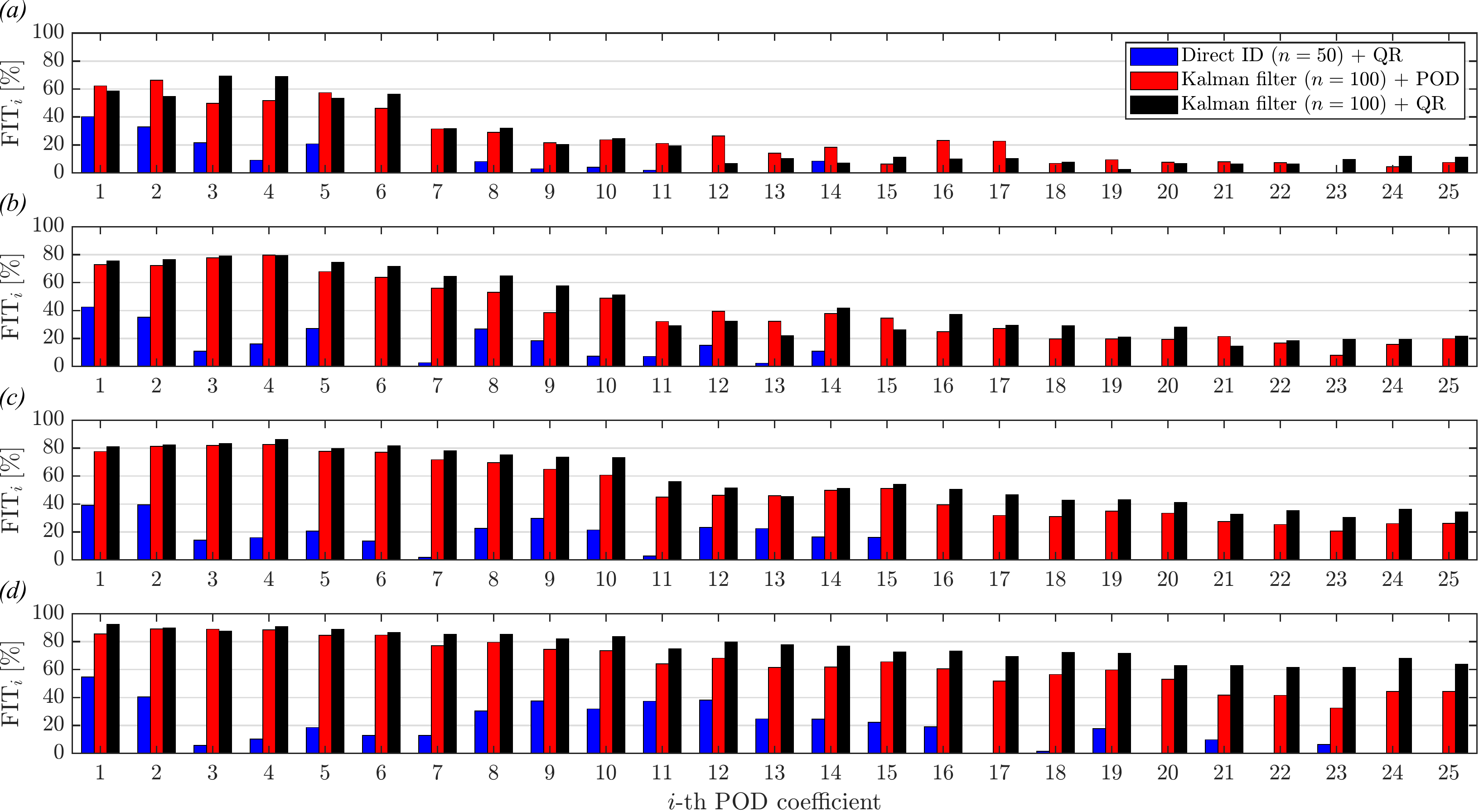}}
    \caption{Histograms of FIT values for the POD temporal coefficients of modes 1-25. Results are shown for $p=4$ ($a$), $p=10$ ($b$), $p=20$ ($c$) and $p=40$ ($d$) measurements. Blue, red and black bars refer to direct identification, Kalman filter with POD peaks and with QR sensors, respectively. Only the validation dataset is used to compute the FIT.}
    \label{fig:FIT POD coefficient}
 \end{figure}

To further assess the effect of $p$, we investigate the reconstruction accuracy of the POD temporal coefficients which, ultimately, indicates how well the model has been trained to reproduce the dynamics of the original system. Histograms of the FIT, defined as $\textup{FIT}_i\: [\%] = 100 \left (1 -  \frac{   \overline{\left(a_i(t)-\hat{a}_i(t)\right)^2} }{\overline{a_{i}^{2}(t)}} \right )$, are plotted for POD modes 1-25 in figure \ref{fig:FIT POD coefficient}. With $p=4$ (top row), the most energetic modes (up to the $6^{th}$) are estimated more accurately, while for higher modes the FIT is less than 20\%. The Kalman filter models are the most accurate, while QR sensors (black bars) and POD peaks (red bars) show similar trends. Closer inspection reveals that QR sensors allow more accurate identification of the main vortex shedding mode pair (i.e.\ modes 3-4), whilst POD peaks resolve the first mode pair (associated with the gap flow dynamics) slightly better. Using $p=10$ measurements (second row from top) improves the quality of reconstruction for all modes.  With further increase to $p=20$ and $40$, third and fourth (bottom) rows, the QR sensor placement strategy clearly outperforms the POD strategy for all modes (the improvement is most notable for modes with order higher than about 8). Thus, the slightly higher scalar FIT with QR sensors shown in figure \ref{fig:FIT curves} arises from the significant improvement in the reconstruction of the higher order modes that are localised further downstream in the wake (and because these modes have lower energy content, the improvement in the scalar FIT is modest). 

\subsection{Reconstruction of instantaneous flow \label{subsubsec:Reconstruction of instantaneous flow}}
\begin{figure} [ht]
    \centerline{\includegraphics[width=\textwidth]{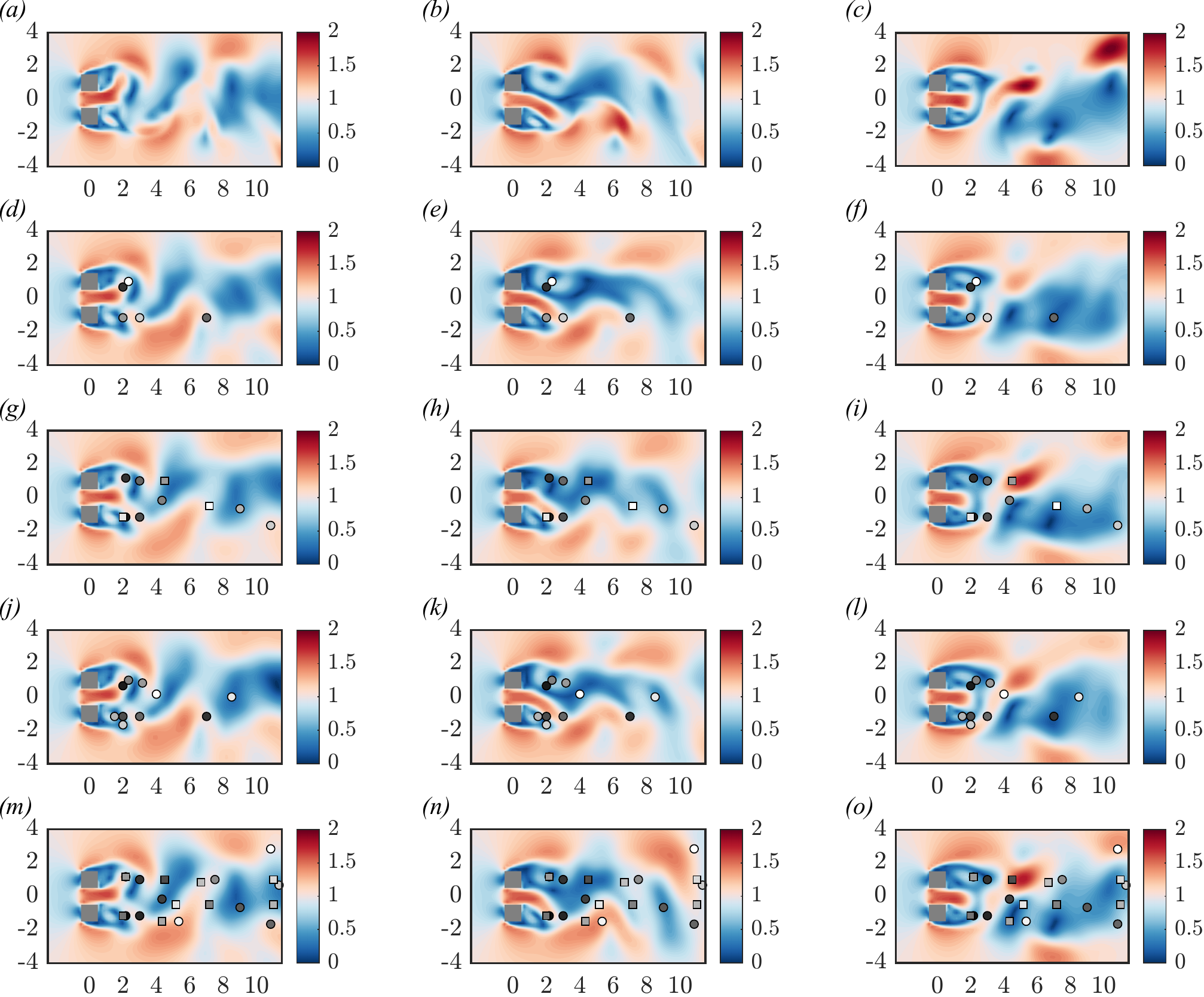}}
    \caption{Contours of velocity magnitude at $t=2000$ (left column), $t=2012$ (middle column) and $t=2024$ (right column) obtained from Kalman filter model of order $n=100$. DNS data ($a$-$c$), reconstruction with 10 POD sensor measurements ($d$-$f$), 10 QR sensor measurements ($g$-$i$), 20 POD sensor measurements ($j$-$l$) and 20 QR sensor measurements ($m$-$o$). Square $(\square)$ and circular $(\bigcirc)$ markers indicate $u'$ and $v'$ measurements, respectively. POD peaks are indicated with circular markers, but both velocity components are recorded.} 
    \label{fig:instantaneous velocity magnitude reconstruction}
\end{figure}

The fluctuations are reconstructed from  equation \eqref{eq: truncated_reconstruction} using the estimated temporal coefficients, $\hat{a}_i$, and the mean flow is added to obtain the instantaneous velocity, $u=\overline{u}+u'$ and $v=\overline{v}+v'$. Figure  \ref{fig:instantaneous velocity magnitude reconstruction} shows contours of the instantaneous velocity magnitude at three time instants $t=2000, 2012 \: \textup{and} \: 2024$ which belong to the validation dataset; results are shown only for the two Kalman filter models with $p=10$ and $p=20$.

\noindent The snapshots from DNS (top row) clearly illustrate the flapping motion of the jet emanating from the gap between the cylinders, the near wake vortices behind each cylinder and the wake merging downstream. We notice that the recirculating vortical regions and the local phase of the jet flow motion are well captured with 10 measurements both from POD, panels (d)-(f), as well as the QR sensors, panels (g)-(i). This is an important achievement for a linear model because the gap flow is characterised by nonlinear dynamics that influences the behaviour of the downstream flow and modulates the aerodynamic coefficients, see figure \ref{fig:aero time history}. The flow field is predicted accurately up to $x\approx6$, in agreement with the fact that 8 out of 10 measurements are extracted in $x<6$. Another interesting result from panels (d)-(i) is that the large convective structures in the mixed wake are predicted quite well, except for the patch of the fast moving fluid of panel (c) at the top right corner that is not reconstructed well with 10 measurements. Overall, there are marginal differences between POD and QR sensors, at least visually. 

A significant improvement in accuracy is seen for 20 QR measurements, panels (m)-(o). Compared to POD sensors, panels (j)-(l), the increased spread of the QR sensors allows for  better sensing of the downstream wake, which is resolved with visibly greater accuracy. The thick wake behind the top cylinder in panel (b) is better captured by the QR sensors compared to POD sensors. This feature is also well reproduced in the reconstruction with 10 measurements. The high-velocity patch in the top right corner of panel (c) is also captured with higher accuracy by the QR sensors. Overall, 20 QR sensors yield exceptionally good prediction of the spatio-temporal dynamics of large-scale vortical structures and achieve relatively good resolution of the smaller scales of the flow.  This is in agreement with the results shown in figure \ref{fig:FIT POD coefficient}, where almost all temporal coefficients have higher FIT values with the QR sensors compared to POD sensors for $p=20$. 

\begin{figure}
    \centerline{\includegraphics[width=\textwidth]{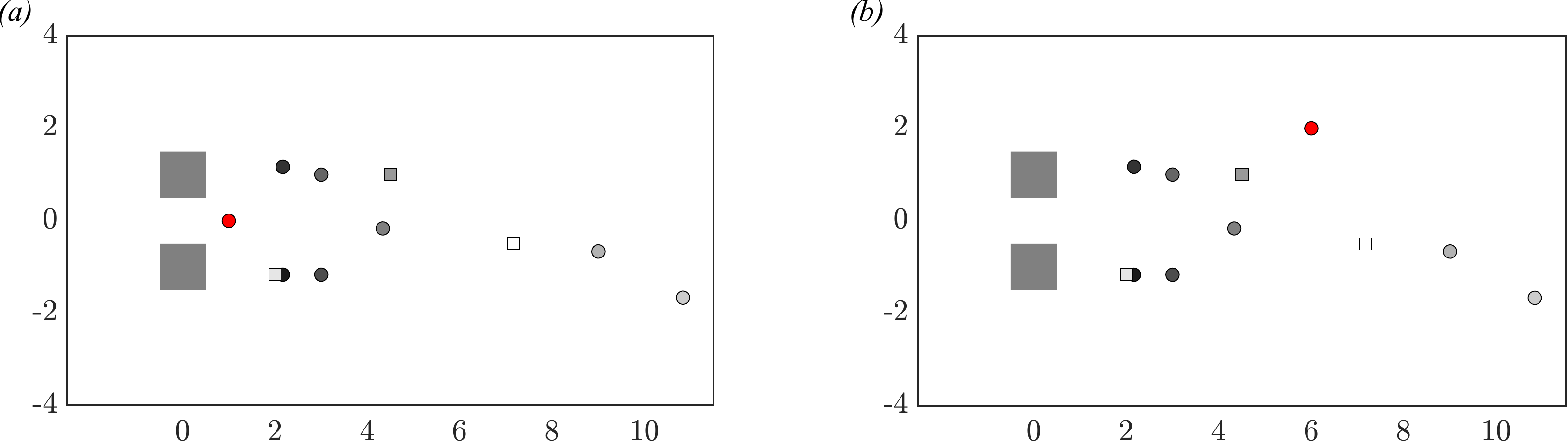}}
    \par
    \vspace{0.3cm}
    \centerline{\includegraphics[width=\textwidth]{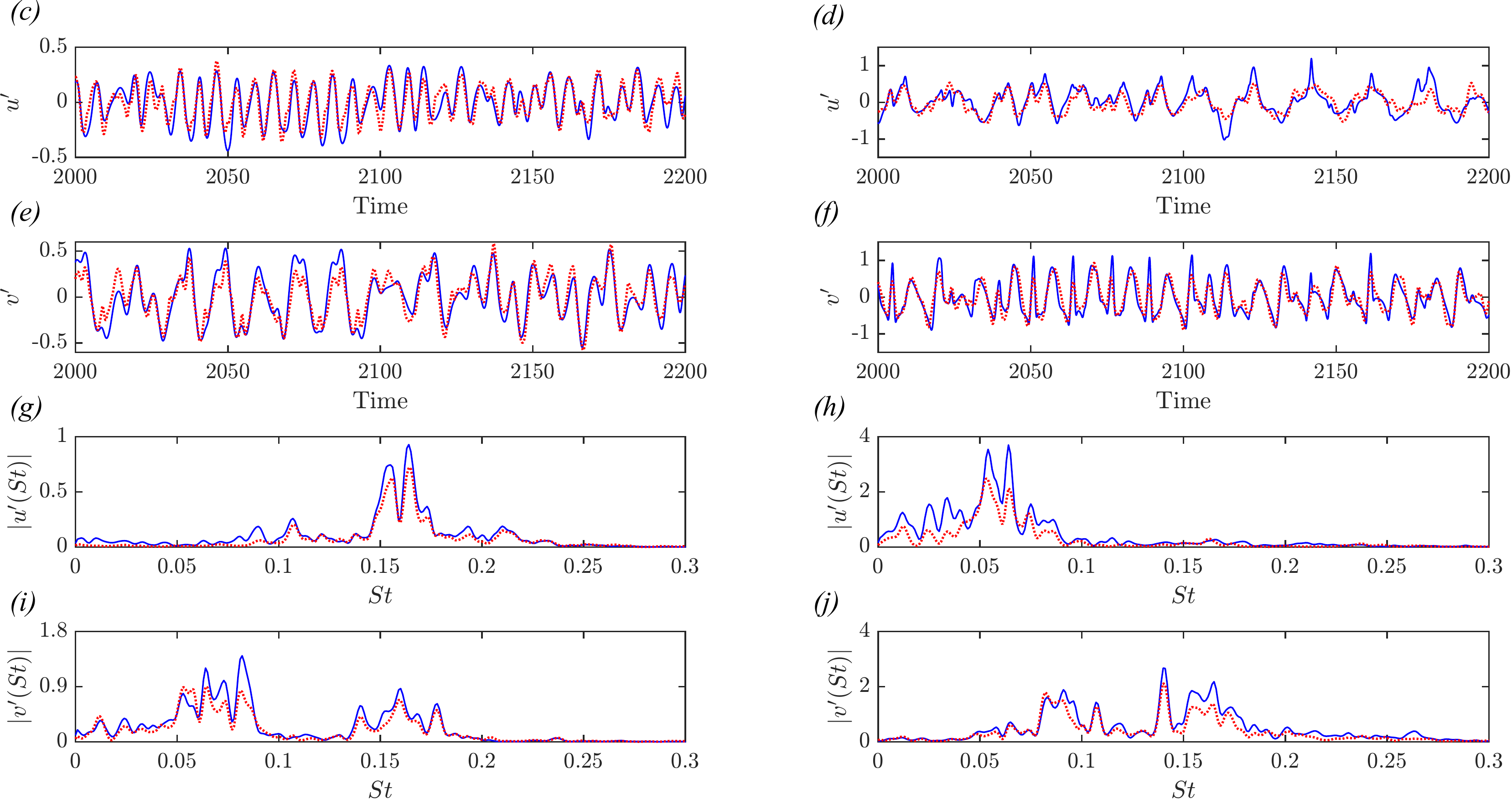}}
    \caption{Reconstructed velocity fluctuations at probe points $(x,y)=(1,0)$ (left column) and $(x,y)=(6,2)$ (right column) from the Kalman filter model of order $n=100$, using $p=10$ QR sensor measurements. Locations of the probe points and sensors ($a$-$b$), temporal reconstruction of  $u'(t)$ ($c$-$d$), $v'(t)$ ($e$-$f$), $u'$ spectra ($g$-$h$) and $v'$ spectra ($i$-$j$). The probe points are marked with a filled red circle. DNS: blue solid line. Reconstruction from the model: red dotted line.}
    \label{fig:reconstructed spectra}
\end{figure}

Finally, we examine the reconstruction of instantaneous velocity fluctuations at two probe points, one located on the centreline at $(1,0)$ and the other at $(6,2)$. The points are marked with filled red circles in the top row of figure \ref{fig:reconstructed spectra}. The locations are chosen so that we can further investigate the accuracy of the model in the gap flow and in the mixed wake regions, which are representative of  different dynamics. Results are presented only for the  Kalman filter model of order $n=100$ with QR sensors. The reconstructed fluctuations in the time and frequency domains with $p=10$ measurements are compared against the DNS data. The velocity signals, panels (c)-(f), show only minor amplitude mismatches, while the phase is predicted very accurately. As a result, the reconstructed frequency spectra shown in panels (g)-(j), are in good agreement with the DNS data. For point $(0,1)$, the moderate-range frequency, $St=0.168$ (in the $u'$ spectra) and the slower temporal scales, $St=0.063,\:0.091$ (in the $v'$ spectra), are captured by the model. Again, this proves that the dynamic estimator is capable of predicting the dynamic behaviour of the flapping jet. At point $(6,2)$, the new temporal scales that arise from the interaction between the fundamental shedding mode and the flapping jet mode are also reproduced well.

\section{Conclusions}\label{sec:conclusions}
The two-dimensional flow around two side-by-side square cylinders was simulated at $Re=200$ and gap ratio $g^{*}=1$. This parameter combination results in a highly irregular, non-periodic pattern, with strong interaction between the flapping jet and the vortex formation and release pattern. The aerodynamic coefficients and velocity signals have spectral content at two frequency bands, a low frequency  ($St=0.025,\:0.053,\:0.063$) and a moderate one  ($St=0.168$). Visualisation of the evolution of the flow revealed the complexity of the wake dynamics.

The high spatial dimensionality of the DNS dataset was reduced by applying POD. The first 50 modes were able to capture about 97\% of the energy and were retained in the reduced-order model approximation. The subspace system identification algorithm \texttt{n4sid} was then  applied in order to extract two different data-driven dynamic estimators. The construction of the first estimator comprised two separate steps (a) identification of the underlying system dynamics and (b) application of Kalman filter. The second estimator was identified directly from the data. The former approach was found to be more robust and computationally efficient; it is also more flexible and allows to include in the estimator design information on the spatial shapes of the POD modes as well as errors due to  model and measurement uncertainties. Two ideas for sparse sensor placement were also explored, placement at POD mode peaks and at points obtained from a greedy QR pivoting algorithm that solves a submatrix volume optimisation problem. 

Systematic performance analysis demonstrated the superiority of the Kalman filter with QR sensors compared to POD peaks, especially in the reconstruction of the far wake. This is because the former type of sensors were more evenly spread in the domain compared to the latter type that were clustered closer to the cylinders. The accuracy of the Kalman filter was also assessed in terms of instantaneous flow reconstruction. Contours and frequency spectra suggested that the flapping dynamics of the gap flow and  the mixed wake can be predicted by the model with only 10 measurements. From the reconstructed velocity field, the pressure can be  extracted (by solving a Poisson equation) and the time histories of the aerodynamic coefficients can be also estimated. 

This challenging test case demonstrated that a high level of complexity can be handled by appropriately designed data-driven linear models. A natural next step would be the extension of the present approach to control, for example to identify the optimal sensor and actuation placement to suppress the irregular vortex shedding. This research direction is left for future work.

\section*{Acknowledgements.} The authors would like to thank Dr. Wynn and Dr. Guzman-I{\~{n}}igo for helpful discussions and feedback on this work.

\section*{Declaration of Interests.} The authors report no conflict of interest.

\section*{Author ORCIDs.}
\orcidicon{0000-0002-2576-0685} F. Savarino \href{https://orcid.org/0000-0002-2576-0685}{https://orcid.org/0000-0002-2576-0685};

\orcidicon{0000-0003-0594-3107} G. Papadakis \href{https://orcid.org/0000-0003-0594-3107}{https://orcid.org/0000-0003-0594-3107}.

\bibliographystyle{jfm}
\bibliography{references}

\begin{thebibliography}{50}
\expandafter\ifx\csname natexlab\endcsname\relax\def\natexlab#1{#1}\fi
\def\au#1{#1} \def\ed#1{#1} \def\yr#1{#1}\def\at#1{#1}\def\jt#1{\textit{#1}}
  \def\bt#1{#1}\def\bvol#1{\textbf{#1}} \def\vol#1{#1} \def\pg#1{#1}
  \def\publ#1{#1}\def\arxiv#1{#1}\def\org#1{#1}\def\st#1{\textit{#1}}

\bibitem[Adrian(1979)]{Adrian_1979}
{\sc \au{Adrian, R.~J.}} \yr{1979}  \at{Conditional eddies in isotropic
  turbulence}.  \jt{Physics of Fluids}  \bvol{22}~(11),  \pg{2065--2070}.

\bibitem[Adrian \& Moin(1988)]{adrian_moin_1988}
{\sc \au{Adrian, R.~J.} \& \au{Moin, P.}} \yr{1988}  \at{Stochastic estimation
  of organized turbulent structure: homogeneous shear flow}.  \jt{J. Fluid
  Mech}  \bvol{190}.

\bibitem[Amaral {\em et~al.\/}(2021)Amaral, Cavalieri, Martini, Jordan \&
  Towne]{Amaral}
{\sc \au{Amaral, Filipe~R.}, \au{Cavalieri, André~V.G.}, \au{Martini,
  Eduardo}, \au{Jordan, Peter} \& \au{Towne, Aaron}} \yr{2021}
  \at{Resolvent-based estimation of turbulent channel flow using wall
  measurements}.  \jt{Journal of Fluid Mechanics}  \bvol{927},  \pg{A17}.

\bibitem[Anderson \& Moore(1979)]{Anderson_and_moore}
{\sc \au{Anderson, B. D.~O.} \& \au{Moore, J.~B.}} \yr{1979} {\em Optimal
  Filtering\/}.  \publ{Prentice-Hall, Inc.}

\bibitem[Ansys(2009)]{PISO}
{\sc \au{Ansys}} \yr{2009} {\em {FLUENT} 12.0 {Theory Guide}\/}.  \publ{Ansys
  Inc.}

\bibitem[Antoulas(2005)]{Antoulas_2005}
{\sc \au{Antoulas, A.}} \yr{2005} {\em Approximation of Large-Scale Dynamical
  Systems\/}.  \publ{Society for Industrial and Applied Mathematics}.

\bibitem[Bhattacharjee {\em et~al.\/}(2020)Bhattacharjee, Klose, Jacobs \&
  Hemati]{Bhattacharjee}
{\sc \au{Bhattacharjee, D.}, \au{Klose, B.}, \au{Jacobs, G.~B.} \& \au{Hemati,
  M.~S.}} \yr{2020}  \at{Data-driven selection of actuators for optimal control
  of airfoil separation}.  \jt{Theoretical and Computational Fluid Dynamics}
  \bvol{34}~(4),  \pg{557--575}.

\bibitem[Brunton \& Noack(2015)]{Brunton_Noack_2015}
{\sc \au{Brunton, S.L.} \& \au{Noack, B.R.}} \yr{2015}  \at{{Closed-Loop
  Turbulence Control: Progress and Challenges}}.  \jt{Applied Mechanics
  Reviews}  \bvol{67}~(5), 050801.

\bibitem[Brunton \& Kutz(2019)]{Bnew}
{\sc \au{Brunton, S.~L.} \& \au{Kutz, J.~N.}} \yr{2019} {\em Data-Driven
  Science and Engineering: Machine Learning, Dynamical Systems, and Control\/}.
   \publ{Cambridge University Press}.

\bibitem[Burattini \& Agrawal(2013)]{Burattini}
{\sc \au{Burattini, P.} \& \au{Agrawal, A.}} \yr{2013}  \at{Wake interaction
  between two side-by-side square cylinders in channel flow}.  \jt{Computers
  and Fluids}  \bvol{77},  \pg{134--142}.

\bibitem[Callaham {\em et~al.\/}(2019)Callaham, Maeda \&
  Brunton]{Callaham_et_al_2019}
{\sc \au{Callaham, J.L.}, \au{Maeda, K.} \& \au{Brunton, S.L.}} \yr{2019}
  \at{Robust flow reconstruction from limited measurements via sparse
  representation}.  \jt{Phys. Rev. Fluids}  \bvol{4},  \pg{103907}.

\bibitem[Carter {\em et~al.\/}(2021)Carter, De~Voogt, Soares \&
  Ganapathisubramani]{Carter}
{\sc \au{Carter, D.W.}, \au{De~Voogt, F.}, \au{Soares, R.} \&
  \au{Ganapathisubramani, B.}} \yr{2021}  \at{Data-driven sparse reconstruction
  of flow over a stalled aerofoil using experimental data}.  \jt{Data-Centric
  Engineering}  \bvol{2},  \pg{e5}.

\bibitem[Chen \& Rowley(2011)]{Chen_Rowley_2011}
{\sc \au{Chen, KK.} \& \au{Rowley, C.W.}} \yr{2011}  \at{$h_2$ optimal actuator
  and sensor placement in the linearised complex ginzburg–landau system}.
  \jt{Journal of Fluid Mechanics}  \bvol{681},  \pg{241–260}.

\bibitem[Fukami {\em et~al.\/}(2021)Fukami, Fukagata \& Taira]{Fukami}
{\sc \au{Fukami, K.}, \au{Fukagata, K.} \& \au{Taira, K.}} \yr{2021}
  \at{Machine-learning-based spatio-temporal super resolution reconstruction of
  turbulent flows}.  \jt{Journal of Fluid Mechanics}  \bvol{909},  \pg{A9}.

\bibitem[Gera {\em et~al.\/}(2010)Gera, Sharma \& Singh]{Gera}
{\sc \au{Gera, B.}, \au{Sharma, P.~K.} \& \au{Singh, R.~K.}} \yr{2010}
  \at{{CFD} analysis of {2D} unsteady flow around a square cylinder}.
  \jt{International Journal of Applied Engineering Research}  \bvol{1}~(3),
  \pg{602 -- 610}.

\bibitem[Giannopoulos \& Aider(2020)]{Giannopoulos}
{\sc \au{Giannopoulos, R.} \& \au{Aider, J.}} \yr{2020}  \at{Data-driven order
  reduction and velocity field reconstruction using neural networks: The case
  of a turbulent boundary layer}.  \jt{Physics of Fluids}  \bvol{32},
  \pg{095117}.

\bibitem[Green \& Limebeer(1995)]{GreenLimebeer95}
{\sc \au{Green, M.} \& \au{Limebeer, D.J.N.}} \yr{1995} {\em Linear Robust
  Control\/}.  \publ{Prentice Hall, Englewood Cliffs}.

\bibitem[Guastoni {\em et~al.\/}(2021)Guastoni, {Güemes}, Ianiro, Discetti,
  Schlatter, Azizpour \& Vinuesa]{Guastoni_et_al_2021}
{\sc \au{Guastoni, L.}, \au{{Güemes}, A.}, \au{Ianiro, A.}, \au{Discetti, S.},
  \au{Schlatter, P.}, \au{Azizpour, H.} \& \au{Vinuesa, R.}} \yr{2021}
  \at{Convolutional-network models to predict wall-bounded turbulence from wall
  quantities}.  \jt{Journal of Fluid Mechanics}  \bvol{928},  \pg{A27}.

\bibitem[Guezennec(1989)]{guezennec89}
{\sc \au{Guezennec, Y.G.}} \yr{1989}  \at{Stochastic estimation of coherent
  structures in turbulent boundary layers}.  \jt{Phys. Fluids}  \bvol{1},
  \pg{1054--1060}.

\bibitem[Gupta {\em et~al.\/}(2021)Gupta, Madhusudanan, Wan, Illingworth \&
  Juniper]{Juniper}
{\sc \au{Gupta, V.}, \au{Madhusudanan, A.}, \au{Wan, M.}, \au{Illingworth, S.}
  \& \au{Juniper, M.}} \yr{2021}  \at{Linear-model-based estimation in wall
  turbulence: Improved stochastic forcing and eddy viscosity terms}.
  \jt{Journal of Fluid Mechanics}  \bvol{925}.

\bibitem[Guzm\'an-I\~nigo {\em et~al.\/}(2019)Guzm\'an-I\~nigo, Sodar \&
  Papadakis]{Inigo_Sodar_Papadakis_2019}
{\sc \au{Guzm\'an-I\~nigo, J.}, \au{Sodar, M.~A.} \& \au{Papadakis, G.}}
  \yr{2019}  \at{Data-based, reduced-order, dynamic estimator for
  reconstruction of nonlinear flows exhibiting limit-cycle oscillations}.
  \jt{Phys. Rev. Fluids}  \bvol{4},  \pg{114703}.

\bibitem[Guzm{\'a}n-I{\~n}igo {\em et~al.\/}(2014)Guzm{\'a}n-I{\~n}igo, Sipp \&
  Schmid]{guzman2014dynamic}
{\sc \au{Guzm{\'a}n-I{\~n}igo, J.}, \au{Sipp, D.} \& \au{Schmid, P.~J.}}
  \yr{2014}  \at{A dynamic observer to capture and control perturbation energy
  in noise amplifiers}.  \jt{J. Fluid Mech}  \bvol{758},  \pg{728--753}.

\bibitem[Holmes {\em et~al.\/}(1996)Holmes, Lumley \& Berkooz]{H}
{\sc \au{Holmes, P.}, \au{Lumley, J.~L.} \& \au{Berkooz, G.}} \yr{1996} {\em
  Turbulence, Coherent Structures, Dynamical Systems and Symmetry\/}.
  \publ{Cambridge University Press}.

\bibitem[Holmes {\em et~al.\/}(2012)Holmes, Lumley, Berkooz \&
  Rowley]{Holmes2012TurbulenceSymmetry}
{\sc \au{Holmes, Philip}, \au{Lumley, John~L.}, \au{Berkooz, Gahl} \&
  \au{Rowley, Clarence~W.}} \yr{2012} {\em {Turbulence, Coherent Structures,
  Dynamical Systems and Symmetry}\/}, 2nd edn.  \publ{Cambridge: Cambridge
  University Press}.

\bibitem[Kailath {\em et~al.\/}(2000)Kailath, Hassibi \&
  Sayed]{Kailath_Hassibi_Sayed_2000}
{\sc \au{Kailath, T.}, \au{Hassibi, B.} \& \au{Sayed, A.~H.}} \yr{2000} {\em
  Linear estimation\/}.  \publ{Prentice-Hall International}.

\bibitem[Kikuchi {\em et~al.\/}(2015)Kikuchi, Misaka \&
  Obayashi]{Kikuchi_et_al_2015}
{\sc \au{Kikuchi, R.}, \au{Misaka, T.} \& \au{Obayashi, S.}} \yr{2015}
  \at{Assessment of probability density function based on {POD} reduced-order
  model for ensemble-based data assimilation}.  \jt{Fluid Dynamics Research}
  \bvol{47}~(5),  \pg{051403}.

\bibitem[Kim {\em et~al.\/}(2021)Kim, Kim, Won \& Lee]{Kim_kim_won_lee_2021}
{\sc \au{Kim, H.}, \au{Kim, J.}, \au{Won, S.} \& \au{Lee, C.}} \yr{2021}
  \at{Unsupervised deep learning for super-resolution reconstruction of
  turbulence}.  \jt{Journal of Fluid Mechanics}  \bvol{910},  \pg{A29}.

\bibitem[Ljung(1999)]{ljung1999system}
{\sc \au{Ljung, L.}} \yr{1999} {\em System identification: theory for the
  user\/}.  \publ{Prentice-Hall PTR}.

\bibitem[Loiseau {\em et~al.\/}(2018)Loiseau, Noack \& Brunton]{Loiseau}
{\sc \au{Loiseau, J.~C.}, \au{Noack, B.~R.} \& \au{Brunton, S.~L.}} \yr{2018}
  \at{Sparse reduced-order modelling: Sensor-based dynamics to full-state
  estimation}.  \jt{Journal of Fluid Mechanics}  \bvol{844},  \pg{459--490}.

\bibitem[Ma {\em et~al.\/}(2017)Ma, Kang, Lim, Wu \& Tutty]{Ma}
{\sc \au{Ma, S.}, \au{Kang, C.~W.}, \au{Lim, T. B.~A.}, \au{Wu, C.~H.} \&
  \au{Tutty, O.}} \yr{2017}  \at{Wake of two side-by-side square cylinders at
  low reynolds numbers}.  \jt{Physics of Fluids}  \bvol{29}~(3).

\bibitem[Manohar {\em et~al.\/}(2018)Manohar, Brunton, Kutz \&
  Brunton]{Manohar2}
{\sc \au{Manohar, K.}, \au{Brunton, B.~W.}, \au{Kutz, J.~N.} \& \au{Brunton,
  S.~L.}} \yr{2018}  \at{Data-driven sparse sensor placement for
  reconstruction}.  \jt{IEEE Control Systems}  \bvol{38}~(3),  \pg{63--86}.

\bibitem[Manohar {\em et~al.\/}(2021)Manohar, Kutz \&
  Brunton]{Manohar_et_al_2021}
{\sc \au{Manohar, Krithika}, \au{Kutz, J.~Nathan} \& \au{Brunton, Steven~L.}}
  \yr{2021}  \at{Optimal sensor and actuator selection using balanced model
  reduction}.  \jt{IEEE Transactions on Automatic Control}  \pg{pp. 1--8}.

\bibitem[Martini {\em et~al.\/}(2020)Martini, Cavalieri, Jordan, Towne \&
  Lesshafft]{Martini_Cavalieri_Jordan_Towne_Lesshafft_2020}
{\sc \au{Martini, Eduardo}, \au{Cavalieri, André V.~G.}, \au{Jordan, Peter},
  \au{Towne, Aaron} \& \au{Lesshafft, Lutz}} \yr{2020}  \at{Resolvent-based
  optimal estimation of transitional and turbulent flows}.  \jt{Journal of
  Fluid Mechanics}  \bvol{900},  \pg{A2}.

\bibitem[Martini {\em et~al.\/}(2022)Martini, Jung, Cavalieri, Jordan \&
  Towne]{Martini_Jung_Cavalieri_Jordan_Towne_2022}
{\sc \au{Martini, Eduardo}, \au{Jung, Junoh}, \au{Cavalieri, André~V.G.},
  \au{Jordan, Peter} \& \au{Towne, Aaron}} \yr{2022}  \at{Resolvent-based tools
  for optimal estimation and control via the wiener–hopf formalism}.
  \jt{Journal of Fluid Mechanics}  \bvol{937},  \pg{A19}.

\bibitem[Mikhaylov {\em et~al.\/}(2021)Mikhaylov, Rigopoulos \&
  Papadakis]{Mikhaylov_et_al_2021}
{\sc \au{Mikhaylov, Kirill}, \au{Rigopoulos, Stelios} \& \au{Papadakis,
  George}} \yr{2021}  \at{Reconstruction of large-scale flow structures in a
  stirred tank from limited sensor data}.  \jt{AIChE Journal}  \pg{p. e17348}.

\bibitem[Nair \& Goza(2020)]{Nair_Goza_2020}
{\sc \au{Nair, N.J.} \& \au{Goza, A.}} \yr{2020}  \at{Leveraging reduced-order
  models for state estimation using deep learning}.  \jt{Journal of Fluid
  Mechanics}  \bvol{897},  \pg{R1}.

\bibitem[Oehler \& Illingworth(2018)]{Oehler_Illingworth_2018}
{\sc \au{Oehler, S.F.} \& \au{Illingworth, S.J.}} \yr{2018}  \at{Sensor and
  actuator placement trade-offs for a linear model of spatially developing
  flows}.  \jt{Journal of Fluid Mechanics}  \bvol{854},  \pg{34–55}.

\bibitem[Ooi {\em et~al.\/}(2021)Ooi, Le, Dao, Nguyen, Nguyen \& Ba]{Ooi}
{\sc \au{Ooi, C.}, \au{Le, Q.~Tuyen}, \au{Dao, My~Ha}, \au{Nguyen, Van~Bo},
  \au{Nguyen, Hoang~Huy} \& \au{Ba, Te}} \yr{2021}  \at{Modeling transient
  fluid simulations with proper orthogonal decomposition and machine learning}.
   \jt{International Journal for Numerical Methods in Fluids}  \bvol{93}~(2),
  \pg{396--410}.

\bibitem[Overschee \& Moor(1994)]{Overscheereport}
{\sc \au{Overschee, P.~Van} \& \au{Moor, B.~De}} \yr{1994}  \at{{N4SID}:
  Subspace algorithms for the identification of combined
  deterministic-stochastic systems}.  \jt{Automatica}  \bvol{30}~(1),
  \pg{75--93}.

\bibitem[Overschee \& de~Moor(1996)]{O}
{\sc \au{Overschee, P.~Van} \& \au{de~Moor, B.~L.}} \yr{1996} {\em Subspace
  Identification for Linear Systems: Theory — Implementation —
  Applications\/}.  \publ{Springer US}.

\bibitem[Qin(2006)]{Qin}
{\sc \au{Qin, S.~J.}} \yr{2006}  \at{An overview of subspace identification}.
  \jt{Computers and Chemical Engineering}  \bvol{30}~(10-12),  \pg{1502--1513}.

\bibitem[Rozov \& Breitsamter(2021)]{Rozon_Breitsamter_2021}
{\sc \au{Rozov, V.} \& \au{Breitsamter, C.}} \yr{2021}  \at{Data-driven
  prediction of unsteady pressure distributions based on deep learning}.
  \jt{Journal of Fluids and Structures}  \bvol{104},  \pg{103316}.

\bibitem[Shun \& Chien(2011)]{Shun}
{\sc \au{Shun, Chang~Yen} \& \au{Chien, Ting~Liu}} \yr{2011}  \at{Gap-flow
  patterns behind twin-cylinders at low {Reynolds} number}.  \jt{Journal of
  Mechanical Science and Technology}  \bvol{25}~(11),  \pg{2795--803}.

\bibitem[Sima {\em et~al.\/}(2004)Sima, {Maria Sima} \& {Van
  Huffel}]{Sima_et_al_2004}
{\sc \au{Sima, Vasile}, \au{{Maria Sima}, Diana} \& \au{{Van Huffel}, Sabine}}
  \yr{2004}  \at{High-performance numerical algorithms and software for
  subspace-based linear multivariable system identification}.  \jt{Journal of
  Computational and Applied Mathematics}  \bvol{170}~(2),  \pg{371--397}.

\bibitem[Sipp \& Schmid(2016)]{Sipp_Schmid_2016}
{\sc \au{Sipp, D.} \& \au{Schmid, P.J.}} \yr{2016}  \at{Linear closed-loop
  control of fluid instabilities and noise-induced perturbations: A review of
  approaches and tools1}.  \jt{Applied Mechanics Reviews}  \bvol{68}~(2),
  020801.

\bibitem[Summers {\em et~al.\/}(2016)Summers, Cortesi \&
  Lygeros]{Summers_et_al_2016}
{\sc \au{Summers, T.H.}, \au{Cortesi, F.L.} \& \au{Lygeros, J.}} \yr{2016}
  \at{On submodularity and controllability in complex dynamical networks}.
  \jt{IEEE Transactions on Control of Network Systems}  \bvol{3}~(1),
  \pg{91--101}.

\bibitem[Symon {\em et~al.\/}(2020)Symon, Sipp, Schmid \& McKeon]{Symon}
{\sc \au{Symon, Sean}, \au{Sipp, Denis}, \au{Schmid, Peter~J.} \& \au{McKeon,
  Beverley~J.}} \yr{2020}  \at{Mean and unsteady flow reconstruction using
  data-assimilation and resolvent analysis}.  \jt{AIAA Journal}  \bvol{58}~(2),
   \pg{575--588}.

\bibitem[Tu {\em et~al.\/}(2013)Tu, Griffin, Hart, Rowley, Cattafesta \&
  Ukeiley]{Tu_et_al_2013}
{\sc \au{Tu, Jonathan~H.}, \au{Griffin, John}, \au{Hart, Adam}, \au{Rowley,
  Clarence~W.}, \au{Cattafesta, Louis~N.} \& \au{Ukeiley, Lawrence~S.}}
  \yr{2013}  \at{Integration of non-time-resolved piv and time-resolved
  velocity point sensors for dynamic estimation of velocity fields}.
  \jt{Experiments in Fluids}  \bvol{54}.

\bibitem[Tzoumas {\em et~al.\/}(2016)Tzoumas, Jadbabaie \&
  Pappas]{Tzoumas_et_al_2016}
{\sc \au{Tzoumas, V.}, \au{Jadbabaie, A.} \& \au{Pappas, G.~J.}} \yr{2016}
  Sensor placement for optimal {Kalman} filtering: Fundamental limits,
  submodularity, and algorithms.  \bt{In {\em 2016 American Control Conference
  (ACC)\/}},  \pg{pp. 191--196}.

\bibitem[Yildirim {\em et~al.\/}(2009)Yildirim, Chryssostomidis \&
  Karniadakis]{Yildirim}
{\sc \au{Yildirim, B.}, \au{Chryssostomidis, C.} \& \au{Karniadakis, G.~E.}}
  \yr{2009}  \at{Efficient sensor placement for ocean measurements using
  low-dimensional concepts}.  \jt{Ocean Modelling}  \bvol{27}~(3-4),
  \pg{160--173}.

\end{thebibliography}

\end{document}